\documentclass[aps,a4paper,prl,reprint,superscriptaddress,notitlepage,floatfix,nolongbibliography]{revtex4-2}
\usepackage{amsmath}
\usepackage{amssymb}
\usepackage{graphicx}
\usepackage{bm}
\usepackage{xcolor}
\usepackage{csquotes}
\usepackage[T1]{fontenc}
\usepackage[utf8]{inputenc}
\usepackage{standalone}

\makeatletter
\let\@orig@make@capt@title\@make@capt@title
\def\@make@capt@title#1#2{\@orig@make@capt@title{{\bf #1}}{#2}}
\makeatother

\usepackage{hyperref}
\definecolor{tab_blue}{HTML}{1F77B4}
\hypersetup{%
	breaklinks=true,
	colorlinks=true,
	linkcolor=tab_blue,
	filecolor=tab_blue,
	urlcolor=tab_blue,
	citecolor=tab_blue,
}

\def\odd{\text{odd}}

\newcommand{\rarrow}[1][0.15cm]{\mathrel{%
		\hbox{\rule[\dimexpr\fontdimen22\textfont2-.2pt\relax]{#1}{.4pt}}%
		\mkern-4mu\hbox{\usefont{U}{lasy}{m}{n}\symbol{41}}}}

\newcommand{\larrow}[1][0.15cm]{\mathrel{%
		\hbox{\usefont{U}{lasy}{m}{n}\symbol{40}}
		\mkern-4mu\hbox{\rule[\dimexpr\fontdimen22\textfont2-.2pt\relax]{#1}{.4pt}}%
}}

\makeatother

\begin{document}
	\title{Odd Viscosity Suppresses Intermittency in Direct Turbulent Cascades}

	\author{Sihan Chen}
	\thanks{Both authors contributed equally to this work.}
	\affiliation{Kadanoff Center for Theoretical Physics, The University of Chicago, Chicago, IL 60637, USA}
	\affiliation{James Franck Institute, The University of Chicago, Chicago, IL 60637, USA}
	\author{Xander M. de Wit}
	\thanks{Both authors contributed equally to this work.}
	\affiliation{Department of Applied Physics and Science Education, Eindhoven University of Technology, 5600 MB Eindhoven, Netherlands}
	\author{Michel Fruchart}
	\affiliation{Gulliver, ESPCI Paris, Université PSL, CNRS, 75005 Paris, France}
	\author{Federico Toschi}
	\email{f.toschi@tue.nl}
	\affiliation{Department of Applied Physics and Science Education, Eindhoven University of Technology, 5600 MB Eindhoven, Netherlands}
	\affiliation{CNR-IAC, I-00185 Rome, Italy}
	\author{Vincenzo Vitelli}
	\email{vitelli@uchicago.edu}
	\affiliation{Kadanoff Center for Theoretical Physics, The University of Chicago, Chicago, IL 60637, USA}
	\affiliation{James Franck Institute, The University of Chicago, Chicago, IL 60637, USA}
	\date{\today}

	\begin{abstract}
		Intermittency refers to the broken self-similarity of turbulent flows caused by anomalous spatio-temporal fluctuations. 
		In this Letter, we ask how intermittency is affected by a non-dissipative viscosity, known as odd viscosity (also Hall or gyro-viscosity), which appears in parity-breaking fluids such as magnetized polyatomic gases, 
		electron fluids under magnetic field 
		and spinning colloids or grains. Using a combination of Navier-Stokes simulations and theory, we show that intermittency is suppressed by odd viscosity at small scales. This effect is caused by parity-breaking waves, induced by odd viscosity, that break the multiple scale invariances of the Navier-Stokes equations. Building on this insight, we construct a two-channel helical shell model that reproduces the basic phenomenology of turbulent odd-viscous fluids including the suppression of anomalous scaling. Our findings illustrate how a fully developed direct cascade that is entirely self-similar can emerge below a tunable length scale, paving the way for designing turbulent flows with adjustable levels of intermittency.
	\end{abstract}
	
	\maketitle
	
\begin{figure*}[]
		\centering
		\hspace{-2em}\includegraphics[width = 2.\columnwidth]{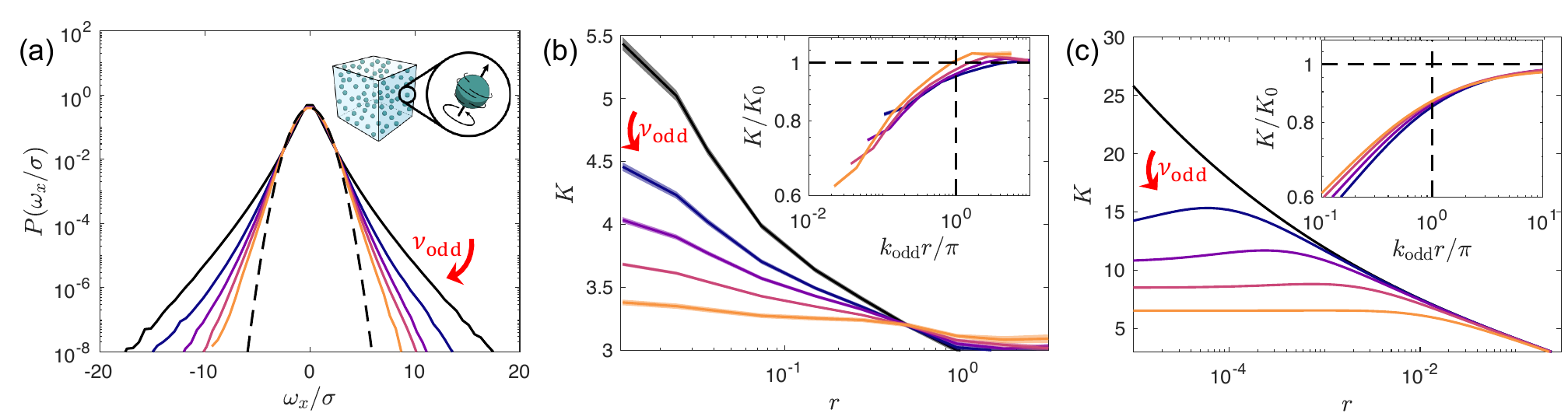}
		\caption{
			\textbf{Intermittency in odd turbulence.}
			(a) Probability distribution of $x$-direction vorticity in DNS (renormalized by the standard deviation), for $\nu_{\rm odd}=0, 3\times 10^{-4}, 6\times 10^{-4}, 1.2\times 10^{-3}, 2.4\times 10^{-3}$. Gaussian distribution is shown in dashed line for comparison. 
			Inset: illustration of a chiral active fluid. Odd viscosity can be caused by self-spinning of particles in $z$ direction.
			(b) The growth of kurtosis $K$ in DNS is suppressed by odd viscosity at small scales, for same $\nu_{\rm odd}$ as in (a). Errorbar is shown as shaded area. (inset) Rescaled kurtosis $K/K_0$ ($K_0$ being the kurtosis for $\nu_{\rm odd}=0$) vs rescaled lengthscale $k_{\rm odd} r/\pi$, with $k_{\rm odd}=\epsilon^{1/4}\nu_{\rm odd}^{-3/4}$. (c) Kurtosis predicted from the modified Parisi-Frisch theory, for $\nu_{\rm odd}=0, 10^{-6},10^{-5},10^{-4},10^{-3},10^{-2}$. (inset) Rescaled kurtosis $K/K_0$ vs rescaled lengthscale $k_{\rm odd} r/\pi$. Note that the scale separation in (c) is considerably larger than what is achievable in DNS in (b). See Sec.~II D for results with same scale separation as that in DNS (SI~\cite{Supp}).}
		\label{fig.1}
	\end{figure*}

	In 3D turbulent fluids, energy is transferred from large to small length scales through a process known as a turbulent cascade: 
	large eddies tend to split into smaller and smaller eddies
	~\cite{benzi2023lectures,davidson2015turbulence,galtier2022physics,Falkovich2001,Frisch1995,Sreenivasan1991}. 
	The hierarchical structure of the cascade suggests that it may exhibit a self-similarity relating larger and smaller eddies~\cite{Kolmogorov1941a,Kolmogorov1941b}. 
	It turns out that this self-similarity is only approximate~\cite{Frisch1995}: 
	smaller eddies exhibit an increasing propensity for extreme velocities, characterized by non-gaussian, fat-tailed distributions~\cite{VanAttaPark1972,anselmet1984high,paladin1987anomalous,benzi1993intermittency,boratav1997structures,l2000analytic,la2001fluid,gotoh2002velocity,shen2002longitudinal,benzi2003intermittency,Biferale2003,chevillard2003lagrangian,biferale2004multifractal,biferale2005particle,arneodo2008universal,benzi2010inertial,saw2018universality,iyer2020scaling,zhou2021turbulence,buaria2023saturation,de2024extreme} (see Fig.~\ref{fig.1}a). 
	This phenomenon is known as intermittency \cite{paladin1987anomalous,Frisch1995}. 
	
	The precise conditions that lead a turbulent cascade to produce intermittency are still unknown~\cite{Alexakis2018,benzi2023lectures}. 
	In the direct cascade of the 3D Navier-Stokes equation, intermittency (i.e. non-gaussianity) increases along the forward cascade, a property known as \enquote{anomalous scaling}.
	Reductions of this tendency have been reported in magnetohydrodynamics~\cite{david2024monofractality}, nonlinear viscosity fluids~\cite{Buzzicotti2020}, fluids on fractal Fourier sets~\cite{Frisch2012,Buzzicotti2016,Lanotte2015,ray2018non} or subject to rotation~\cite{Seiwert2008,Mininni2009,Baroud2003,Muller2007,Rathor2020,Biferale2016}. 
	It has been conjectured that self-similarity may be restored in a direct cascade by a mathematical \enquote{surgery} of Navier-Stokes equations that reduces the backward energy transfer through certain triadic interactions classified by the helicity of the modes involved~\cite{biferale2012inverse}.
	However, we are not aware of any example of a fully developed direct cascade with no intermittency described by the 3D Navier-Stokes equations.

	In this Letter, we show that the increase of intermittency, that typically occurs along a forward turbulent cascade, can be suppressed in the inertial range by parity-breaking waves occurring in fluids that display a non-dissipative viscosity, known as odd viscosity, such as magnetized polyatomic gases, graphene under magnetic field and spinning colloids or grains~\cite{deGroot1962,Hulsman1970,avron1998odd,banerjee2017odd,berdyugin2019measuring,soni2019odd,souslov2019topological,souslov2020anisotropic,han2021fluctuating,markovich2021odd,khain2022stokes,lou2022odd,fruchart2022odd,fruchart2023odd,khain2023trading,musser2023observable}.
	By controlling odd viscosity, we can tune the length scale below which the growth of intermittency terminates, ultimately creating a fully developed forward turbulent cascade that is completely self-similar, i.e. anomalous scaling is suppressed.
	

	{\it Turbulence with odd viscosity}---%
	While the viscosity tensor $\eta_{ijkl}$ is typically assumed to be symmetric ($\eta_{ijkl}=\eta_{klij}$), an antisymmetric part of the viscosity tensor ($\eta^A_{ijkl}=-\eta^A_{klij}$), known as odd, Hall or gyro viscosity, naturally arises in fluids that break both parity and time-reversal symmetry~\cite{fruchart2022odd}. Odd viscosity with rotational symmetry in $\hat{\bm e}_z$ (e.g., generated by self-spinning particles, see Fig.~\ref{fig.1}(a)) modifies the incompressible Navier-Stokes equation to (see Supplemenetary Information (SI)~\cite{Supp})
	\begin{equation}
	\begin{aligned}
	D_t \bm u = -\nabla P + \nu \Delta \bm u + \nu_{\rm odd}\hat{\bm e}_z\times \Delta \bm u + \bm f
	\label{e2.1}
	\end{aligned}
	\end{equation}
	where $\bm u({\bm x},t)$ is the velocity, 
	$D_t=\partial_t+\bm u\cdot \nabla$ the convective derivative, $\bm f$ the helicity-free external forcing, $\nu$ the dissipative viscosity, $\hat{\bm e}_z$ the unit vector along $z$, and $\nu_{\rm odd}$ the odd viscosity. 
	The odd viscosity term, that can be seen as a scale-dependent Coriolis force, does not directly dissipate or transfer energy, but affects energy transfer by inducing \enquote{odd waves} with frequency $\omega_{\pm}(\bm k)=\pm \nu_{\rm odd}k_z|\bm k|$ in the fluid (SI)~\cite{Supp}. Odd waves are parity-breaking, i.e. the frequency changes sign when the definite helicity $\pm$ is flipped~\cite{fruchart2023odd}.
	As shown in Ref.~\cite{de2024pattern}, 
	odd waves progressively arrest the forward energy flux starting from the characteristic scale $k_{\rm odd}=\epsilon^{1/4}\nu_{\rm odd}^{-3/4}$ ($\epsilon$ is the energy injection rate) where $\omega_{\pm}(\bm k)$
	is comparable with the eddy turnover rate.
	While a qualitative picture of the energy spectrum of odd turbulence has emerged~\cite{de2024pattern}, the effect of odd viscosity on intermittency is yet to be elucidated.

	{\it Intermittency in odd turbulence}---%
	We first analyze the vorticity distribution obtained from direct numerical simulations (DNS), which captures the intermittency 
	around the dissipation scale. Details of the DNS method are provided in SI~\cite{Supp}.
	Remarkably, odd viscosity 
	leads to more Gaussian-like distributions (compare the different curves in Fig.~\ref{fig.1}a). 
	Intermittency 
	can be captured by the kurtosis $K=S_4/S_2^2$ of the structure functions $S_p( r)=\langle(\delta_r u)^p\rangle$ in which $\delta_r u = [\bm u(\bm x+\bm r)-\bm u(\bm x)]\cdot \hat{\bm r}$ are known as the longitudinal velocity increment.
	Due to the anisotropy induced by odd viscosity, the structure function is evaluated for $\hat{\bm r}$ perpendicular to $\hat{\bm e}_z$. 
	At the injection scale, the flow has a Gaussian kurtosis $K=3$ imposed by the forcing.
	In ordinary turbulence ($\nu_{\rm odd}=0$), the kurtosis increases as energy cascades towards smaller scales (Fig.~\ref{fig.1}b). 
	For finite $\nu_{\rm odd}$, the growth of the kurtosis is suppressed at small scales where odd viscosity is dominant.
	The rescaled data (inset) shows that the transition between the intermittent and non-intermittent
	regimes is located around $r=\pi/ k_{\rm odd}$, where odd viscosity becomes relevant~\cite{de2024pattern}.

	{\it Modified Parisi-Frisch formalism}---%
	To rationalize this suppression of intermittency, we revisit the Parisi-Frisch framework~\cite{ParisiFrisch1985} relating intermittency to the existence of multiple scale invariances in Navier-Stokes equations.
	
	In the absence of odd viscosity, the unforced dissipation-free Navier-Stokes equation is invariant under the scale transformation
	$\bm x\to \lambda \bm x$, 
	$\quad \bm u \to\lambda^h \bm u$,
	$\quad t\to \lambda^{1-h}t$
	for all $h$.
	Kolmogorov's K41 theory \cite{Kolmogorov1941a,Kolmogorov1941b} effectively assumes that the exponent $h=1/3$ is selected, leading to a self-similar solution. 
	In reality, turbulent flows exhibit a statistical distribution $P_r(h) \sim r^{F(h)}$ of scaling exponents $h$, as allowed by the scaling symmetry. This intertwining of different exponents $h$, referred to as multifractality, breaks the self-similarity of the flow statistics at different scales, heralding the emergence of intermittency~\cite{Frisch1995,she1994universal}. 
	
	\begin{figure}[b!]
		\centering
		\hspace{-2em}\includegraphics[width =0.9\columnwidth]{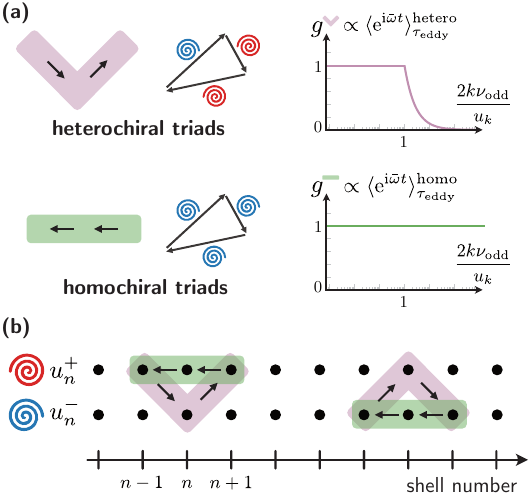}
		\caption{
			\textbf{Effect of odd waves on triadic interactions and helical shell model.}
			(a) Asymmetric effects of odd viscosity on homochiral and heterochiral triads. Odd viscosity strongly affects heterochiral triads by reducing the fraction of resonant triads, $g_k$. Odd viscosity has no effect on local homochiral triads for which $g_k=1$. 
			(b) Two-channel helical shell model. Each shell contains two velocity variables representing modes with $+$ and $-$ chiralities, respectively. Energy transfers forward in heterochiral channel and backward in homochiral channel, through triads formed by nearest neighbors. 
		}
		\label{fig.2}
	\end{figure}

	\begin{figure*}[t]
		\centering
		\hspace{-1em}\includegraphics[width = 2\columnwidth]{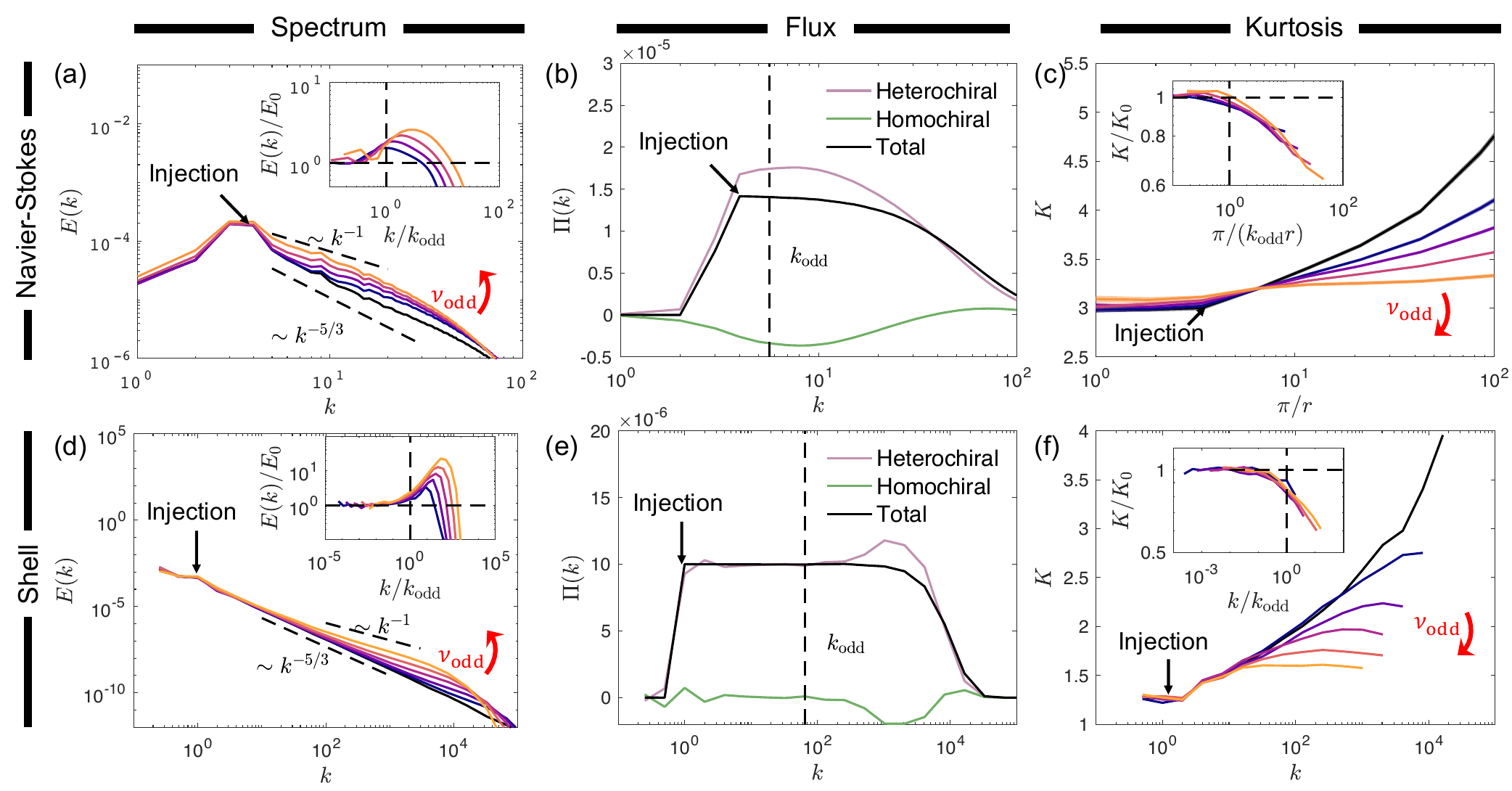}
		\caption{
			\textbf{Odd turbulence in DNS and the helical shell model.}
			(a). Energy spectrum $E(k)$ in DNS, for $\nu_{\rm odd}=0, 3\times 10^{-4}, 6\times 10^{-4}, 1.2\times 10^{-3}, 2.4\times 10^{-3}$. The spectrum in the inertial range changes from a K41 regime ($E(k)\sim k^{-5/3}$) to a wave-affected regime ($E(k)\sim k^{-1}$). (inset) Rescaled spectrum $E(k)/E_0$ ($E_0$ being the spectrum for $\nu_{\rm odd}=0$) vs rescaled wave number $k/k_{\rm odd}$, with $k_{\rm odd}=\epsilon^{1/4}\nu_{\rm odd}^{-3/4}$. (b) Energy flux $\Pi(k)$ and flux loop formed by homochiral and heterochiral components in DNS. $\nu_{\rm odd}=2.4\times 10^{-3}$. (c) Kurtosis in DNS, same as Fig.~\ref{fig.1}(d) with $x$-axis changed from $r$ to $\pi/r$. (inset) Rescaled kurtosis $K/K_0$ ($K_0$ being the kurtosis for $\nu_{\rm odd}=0$) vs rescaled lengthscale $\pi/(k_{\rm odd} r)$. (d) Energy spectrum in the shell model, for $\nu_{\rm odd}=0, 2.6\times 10^{-7}, 1.0\times 10^{-6}, 4.1\times 10^{-6}, 1.6\times 10^{-5}, 6.6\times 10^{-5}$. (inset) Rescaled spectrum $E(k)/E_0$ vs rescaled wave number $k/k_{\rm odd}$. (e) Energy fluxes in the shell model, for $\nu_{\rm odd}=6.6\times 10^{-5}$. (f) Kurtosis $K(k_n)=S_4(k_n)/S_2^2(k_n)$ truncated at the dissipation scale, with $S_p(k_n)=\langle (|u_{n-1}^+| |u_{n}^+| |u_{n+1}^{+}|)^{p/3}\rangle$~\cite{Lvov1998ImprovedSM,de2024extreme}, for $\nu_{\rm odd}$ values same as that in (d). (inset) Rescaled kurtosis $K/K_0$ vs rescaled wavenumber $k/k_{\rm odd} $. In the shell model simulations, we have used $N=30$ shells, $\lambda=2$, $k_0=1/8$, an injection rate $\epsilon=10^{-5}$, $\beta=0.3$, $\gamma=0.2$ and $\nu=10^{-9}$. A constant injection forcing $f_n^{\pm}=0.5\delta_{n,3}\epsilon/u_{n}^{\pm*}$ is used. Note that the scale separation in the shell model (bottom row) is considerably larger than what is achievable in DNS (top row). See Sec.~II D for results with same scale separation as that in DNS (SI~\cite{Supp}).
		}
		\label{fig.3}
	\end{figure*}

	Intermittency is suppressed by odd viscosity because it violates the multiple scaling symmetry of Navier-Stokes equations (SI~\cite{Supp}).
	We conjecture that intermittent solutions $\delta_r u\sim r^h$ with arbitrary $h$ can still appear at large scales where $\nu_{\rm odd}$ has a small effect, and that they reduce to the same wave-affected solution $\delta_r u \sim \nu_{\rm odd}^{1/4}$ at small scales, governed by a single exponent $h=0$ (SI~\cite{Supp}). 

	We model this crossover by adapting the approach of Ref.~\cite{benzi2009fully} 
	and obtain
	\begin{equation}
	\begin{aligned}
	\delta_r u &= r^h\left[1+r_{\rm odd}(h)/r\right]^h\,,\\
	P_r(h) &= r^{F(h)}\left[1+r_{\rm odd}(h)/r\right]^{F(h)}\,.
	\label{e3.2}
	\end{aligned}
	\end{equation}
	in which $r_{\rm odd}(h)=\nu_{\rm odd}^{1/(1+h)}$ is the characteristic lengthscale, which is determined by equating the eddy turnover time $\tau_{\rm eddy}=r/\delta_r u$ and the odd wave period $\tau_{\rm odd}=\nu_{\rm odd}^{-1}r^2$~\cite{de2024pattern}.
	Eq.~(\ref{e3.2}) qualitatively reproduces the 
	arrest of intermittency observed in DNS, see Fig.~\ref{fig.1}c ($F(h)$ is adopted from Ref.~\cite{she1994universal}). 
	For small odd viscosity, $K(r)$ is non-monotonic, in agreement with DNS with extended inertial range (SI~\cite{Supp}). This is due to the variation of $r_{\rm odd}(h)$, such that the crossover between two regimes takes place at different lengthscale for different $h$. Similar non-monotonicity was observed in Ref.~\cite{benzi2009fully}.

	{\it Nonlinear energy transfer}---%
	In the Navier-Stokes equation, nonlinear energy transfer 
	occurs through triads of modes with wavenumbers satisfying $\bm{k} + \bm{p} + \bm{q} = \bm{0}$ \cite{Dar2001,Alexakis2005,Mininni2005,Alexakis2018,verma2019energy}.
	As helicity is conserved, 
	it is convenient to consider modes with definite helicity $s_k = \pm 1$.
	When odd waves are present, these modes get out of phase. 
	As a consequence, the contribution to the non-linear energy transfer of a given triad is multiplied with $e^{i \bar\omega t}$ in which $\bar\omega=\omega_{s_k}(\bm k)+\omega_{s_p}(\bm p)+\omega_{s_q}(\bm q)$~\cite{de2024pattern}.
	When $\bar\omega\tau_{\rm eddy}\gg1$, the fast oscillation effectively cancels the time-averaged energy transfer during $\tau_{\rm eddy}$.
	Therefore, in odd turbulence energy can only transfer through quasi-resonant triads (for which $\bar\omega\tau_{\rm eddy}\leq1$)~\cite{Zakharov1992Kolmogorov}. 
	We estimate the average energy transfer from the fraction $g_k$ of quasi-resonant triads and the energy transfer rate $\epsilon \sim k u_k^3$ in ordinary turbulence, leading to $\epsilon\sim g_k k u_k^3$. 
	
	For quasi local triads ($k=|\bm k|\approx|\bm p|\approx |\bm q|$)~\cite{eyink2009localness} we have $\bar \omega \sim \nu_{\rm odd} k^2$ and $\tau_{\rm eddy}=k^{-1}u_k^{-1}$, hence $g_k=1$ for $\nu_{\rm odd} \ll u_k/k$, corresponding to a regular turbulence regime for $k\ll k_{\odd}$. For $\nu_{\rm odd} \gg u_k/k$ ($k\gg k_{\odd}$), odd waves arrest the energy transfer with $g_k\sim u_k/(k\nu_{\rm odd})$ and $\epsilon\sim u_k^4/\nu_\odd$. Interestingly, odd waves have a strong impact on the forward transfer of heterochiral traids ($+-+$, $-+-$) but only a weak impact on the inverse transfer of homochiral triads ($+++$, $---$)~\cite{biferale2012inverse,plunian2020inverse}, see Fig.~\ref{fig.2}(a), because for homochiral triads with exact locality we always have $\bar \omega=0$ and $g_k=1$~\footnote{For $s_k=s_p=s_q$, $\bar{\omega} = s_k k_z|\bm k|+s_p p_z|\bm p| + s_q q_z|\bm q|=s_k|\bm k|(k_z+p_z+q_z)=0$, because $\bm k $+$\bm p $+$\bm q =0$.}.
	This asymmetry enhances the flux loop formed by the two channels (Fig.~\ref{fig.3}b) and triggers the aforementioned breakdown of the multiple scale invariances of Navier-Stokes equations, which ultimately suppresses intermittency (SI~\cite{Supp}). Importantly, such a mechanism requires the energy to be symmetrically distributed on modes with both definite helicities ($+$ and $-$), which ensures that the energy flux is not dominated by homochiral channels only.

	\begin{figure}[]
		\hspace{-1em}\includegraphics[width = 1\columnwidth]{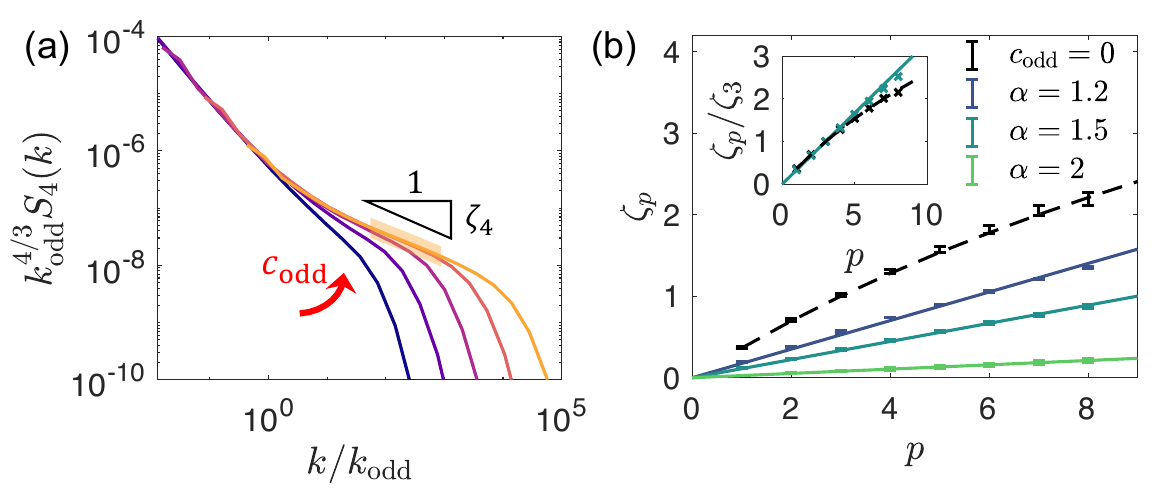}
		\caption{\textbf{Scaling exponents in the helical shell model.} (a) Rescaled structure function $k_{\odd}^{4/3}S_4$ ($k_{\odd}=\epsilon^{1/(3\alpha-2)}c_{\odd}^{3/(2-3\alpha)}$) in shell model with generalized odd viscosity for $c_{\rm odd}/c_0=0, 4^m$, with $c_0 = (\epsilon/2)^{1/2-\alpha/4}\nu^{3\alpha/4-1/2}$, $\alpha=1.5$ and $m=3-7$. (b) Exponents in the wave-affected regime $\zeta_p$ for $\alpha=1.2, 1.5, 2$. Exponents for $\alpha=1.5$ are fitted by the shaded area in (a), see SI for details~\cite{Supp}. Solid lines are monofractal fit $\zeta_p=h_\alpha p$. Error bars are estimated from shifting the fitting range by $\pm1$ shells. Exponents for $c_{\odd}=0$ are shown for comparison, which is consistent with She-Lévêque (dashed line)~\cite{she1994universal}. (inset) $\zeta_p/\zeta_3$ for $c_{\rm odd}=0$ and for $\alpha=1.5$.
			Shell model parameters are same as that in Fig~\ref{fig.3}. Inertial range in the simulation is extended (SI~\cite{Supp}).	}
		\label{fig.4}
	\end{figure}

	{\it Helical shell model}---%
	We now construct a two-channel helical shell model that captures most odd-turbulence phenomenology (Fig.~\ref{fig.2}b)~\footnote{See Ref.~\cite{Kiran2024} for a non-helical single-channel shell model}. Shell models have proven to be a particularly useful tool to study the inertial-range intermittency in turbulence, as they can probe scales far beyond what is accessible in DNS~\cite{Biferale2003} (Although they behave differently from DNS at the dissipation scale~\cite{benzi2009fully}).
	\nocite{Kiran2024}
	Shells are equispaced on a logarithmic scale $k_n=k_0 \lambda^n$ and each shell has two complex dynamical variables, $u_n^+$ and $u_n^-$, representing velocity fluctuations of modes with $+$ and $-$ definite helicities, respectively. We impose conservation of energy $E=\sum_n|u_n^+|^2+\sum_n|u_n^-|^2$ and helicity $H=\sum_nk_n|u_n^+|^2-\sum_nk_n|u_n^-|^2$~\cite{benzi1996helical}, resulting in 
	\begin{equation}
	\begin{aligned}
	\partial_t u_{n}^{\pm}= B_{n}^{\pm,\rarrow} + B_{n}^{\pm,\larrow}+f_{n}^{\pm}-\nu k_n^2 u_{n}^{\pm},
	\label{e5.1}
	\end{aligned}
	\end{equation}
	where $f_n$ is the forcing, and
	\begin{equation*}
	\begin{aligned}
	B^{\pm,\rarrow}_n&={a}^{\pm,\rarrow}_n u^{\pm}_{n+2}u_{n+1}^{\mp*}+ b^{\pm,\rarrow}_nu^{\mp}_{n+1}u_{n-1}^{\mp*}- c^{\pm,\rarrow}_nu^{\mp}_{n-1}u_{n-2}^{\pm} \\
	B^{\pm,\larrow}_n&= a^{\pm,\larrow}_n u^{\pm}_{n+2}u_{n+1}^{\pm*}+ b^{\pm,\larrow}_nu^{\pm}_{n+1}u_{n-1}^{\pm*}- c^{\pm,\larrow}_nu^{\pm}_{n-1}u_{n-2}^{\pm}
	\label{e5.2}
	\end{aligned}
	\end{equation*}
	are the nonlinear transfer of heterochiral and homochiral channels, illustrated in purple and green in Fig.~\ref{fig.2}.
	When ${a}^{\pm,\rarrow}_n ={a}^{\pm,\larrow}_n =k_n$, the channels reduce to the Sabra shell models in the forward and inverse cascade regimes, respectively \cite{gledzer1973hydrodynamic,yamada1988lyapunov,Lvov1998ImprovedSM,gilbert2002inverse,boffetta2011shell} ($b_n$ and $c_n$ are determined from conservation laws, SI~\cite{Supp}). 
	To reproduce the effect of odd viscosity, we let
	\begin{equation*}
	\begin{aligned}
	{a}^{\pm,\rarrow}_n &=k_n g_n^{\pm,\rarrow}=k_n\left(1+\frac{\nu_{\rm odd}k_n}{|u_{n+2}^{\pm}|+|u_{n+1}^{\mp}|+|u_{n}^{\pm}|}\right)^{-1} \\
	{a}^{\pm,\larrow}_n &=\beta k_n g_n^{\pm,\larrow}=\beta k_n\left(1+\frac{\gamma\nu_{\rm odd}k_n}{|u_{n+2}^{\pm}|+|u_{n+1}^{\pm}|+|u_{n}^{\pm}|}\right)^{-1}
	\label{e5.3}
	\end{aligned}
	\end{equation*}
	where $0<\beta<1$ ensures that the model is dominated by the forward cascade when $\nu_{\rm odd}=0$. This functional form is chosen to introduce a transition around $\nu_{\rm odd}\approx |u_n|/k_{n}$. Here, $g_n$ mimics the fraction of resonant triads: $g_n\approx 1$ for small $\nu_{\rm odd}$ and $g_n\sim |u_n|/(k_n\nu_{\rm odd})$ for large $\nu_{\rm odd}$. 
	Finally, we choose $0<\gamma<1$ so that the suppression effect is weaker for the homochiral channel. 
	In the SI, we generalize the model to include \enquote{generalized odd viscosity}, for which $\omega_{\pm}(\bm k) = \pm c_{\rm odd}k_z|\bm k|^{\alpha-1}$~\cite{Supp}.
	
	As shown in Fig.~\ref{fig.3}e, results from the numerical simulation of this shell model reproduce the energy flux observed in Navier-Stokes simulation for $k>k_{\rm odd}$ as well as other qualitative features and scaling relations (Fig.~\ref{fig.3}a, d).
	In the SI~\cite{Supp}, we show that a flux loop state~\cite{Alexakis2018} consistent with DNS results~\cite{de2024pattern} is observed in the inverse cascade regime where forcing is injected at small lengthscale $k_{\rm in}>k_{\rm odd}$. 
	
	{\it Suppression of anomalous scaling}---Our shell model results (Figs.~\ref{fig.3}f) mimic the DNS kurtosis (Figs.~\ref{fig.3}c) and 
	its non-monotonic dependence predicted by the modified Parisi-Frisch argument. We also verify that intermittency is suppressed in higher-order structure functions up to $p=8$, in both DNS and shell model (SI~\cite{Supp}).
	For generalized odd viscosity with $\alpha>2/3$, we find a wave-affected scaling regime of the structure functions for $k\gg k_{\odd}$, where the wave-affected exponents $\zeta_p$ are extracted (see Fig.~\ref{fig.4}(a) for an example).
	Fig.~\ref{fig.4}(b) shows that the $\zeta_p$ extracted in the wave-affected regime are consistent with the self-similar exponents $\zeta_p=h_{\alpha}p$ (solid lines), in contrast with the anomalous exponents usually observed in forward cascades (black dashed line). The $h_\alpha$ we measure are close to the theoretical approximation $h_\alpha\approx1/2-\alpha/4$ (SI~\cite{Supp}).
	

	To sum up, we have demonstrated the existence of a fully developed forward turbulent energy cascade that is entirely self-similar below a tunable length scale, paving the way for designing turbulent flows with adjustable levels of intermittency. In future work it would be interesting to generalize the mechanism studied here to more complex flows where other sources of waves may exist~\cite{zhou2017rayleigh}.
	
	\medskip
	
	\begin{acknowledgments}
		{\it Acknowledgments}---We thank Alexandros Alexakis, Luca Biferale and Tali Khain for discussions.
		We are grateful for the support of the Netherlands Organisation for Scientific Research (NWO) for the use of supercomputer facilities (Snellius) under Grant No. 2021.035, 2023.026. This publication is part of the project “Shaping turbulence with smart particles” with project number OCENW.GROOT.2019.031 of the research programme Open Competitie ENW XL which is (partly) financed by the Dutch Research Council (NWO).
		S.C. and M.F. acknowledge a Kadanoff–Rice fellowship funded by the National Science Foundation under award no. DMR-2011854. M.F. acknowledges partial support from the National Science Foundation under grant DMR-2118415 and the Simons Foundation. V.V. acknowledges partial support from the Army Research Office under grant W911NF-22-2-0109 and W911NF-23-1-0212. M.F. and V.V acknowledge partial support from the France Chicago center through a FACCTS grant. This research was partly supported from the National Science Foundation through the Center for Living Systems (grant no. 2317138), the National Institute for Theory and Mathematics in Biology and the Chan Zuckerberg Foundation. This work was completed in part with resources provided by the University of Chicago’s Research Computing Center.
		
		{\it Data availability}---The code used for simulating the helical shell model, processing the data, generating the figures, as well as the data generated during this study are available on Zenodo \cite{chen_2024_zenodo} under the 2-clause BSD licence.
	\end{acknowledgments}


	\bibliography{citation}

\end{document}


\title{Supplementary Information:\\Odd viscosity suppresses intermittency in direct turbulent cascades}
\maketitle

\section{Odd viscosity}
In this section we introduce odd viscosity and emphasize its non-dissipative nature. 

In fluids, the stress tensor $\sigma_{ij}$ is related to the strain rate via a viscosity tensor $\epsilon_{ijkl}$: $\sigma_{ij}=\eta_{ijkl}\partial_lu_k$. The viscosity tensor is usually assumed to be symmetric, i.e., $\eta_{ijkl}=\eta_{klij}$, which applies to passive fluids. However, in active and parity-breaking fluids, e.g., fluids formed by self-spinning particles, asymmetric viscosity tensor ($\eta_{ijkl}\neq\eta_{klij}$) arises because the system is out of equilibrium~\cite{fruchart2022odd}. We decompose the viscosity tensor into a symmetric part $\eta^S_{ijkl} = (\eta_{ijkl}+\eta_{klij})/2$ and an antisymmetric part $\eta^A_{ijkl} = (\eta_{ijkl}-\eta_{klij})/2$. The incompressible Navier-Stokes equation reads:
\begin{equation}
\begin{aligned}
\rho_0 D_t u_j &= -\nabla_j P + \partial_i(\eta_{ijkl} \partial_lu_k)\\
&= -\nabla_j P + \eta \nabla u_j+\partial_i(\eta^A_{ijkl} \partial_lu_k)\,,
\label{S0.1}
\end{aligned}
\end{equation}
where $\rho_0$ is the density. In the last equality, the second term is the dissipation term that arises from $\eta^S_{ijkl}$.

In this work we consider odd viscosity generated by the self-spinning of particles in $\hat{\bm e}_z$. This corresponds to a cylindrically symmetric viscosity tensor which has eight independent odd viscosities~\cite{khain2022stokes}. Only two independent odd viscosity terms $\eta^o_2$ and $\eta^o_2$ remain in the incompressible Navier-Stokes equation~\cite{de2024pattern}:
\begin{align}
    \label{S0.2}
        \rho_0 D_t \bm{u} &= -\nabla P + \eta \Delta \bm{u}\\ \nonumber
    &+ 
    \eta_1^{\text{o}}
    \begin{bmatrix}
        (\partial_x^2 + \partial_y^2)u_y\\
        \\
        -(\partial_x^2 + \partial_y^2)u_x\\
        \\
        0
    \end{bmatrix} 
    +
     \eta_2^{\text{o}} 
     \begin{bmatrix}
         -\partial_z^2 u_y - \partial_y \partial_z u_z\\
        \\
         \partial_z^2 u_x + \partial_x \partial_z u_z\\
        \\
        \partial_z(\partial_y u_x - \partial_x u_y)
    \end{bmatrix}\\ \nonumber
\end{align}
One example of the corresponding odd viscosity tensor is 
\begin{equation}
\begin{aligned}
\eta^A_{ijkl} = \frac{\eta^o_1}{2}(\delta_{ij}\delta_{ik}\epsilon_{il3} + \delta_{ij}\delta_{il}\epsilon_{ik3}- \delta_{kl}\delta_{ki}\epsilon_{kj3}-\delta_{kl}\delta_{kj}\epsilon_{ki3})
+ \frac{\eta^o_2}{2}(\delta_{j3}\delta_{l3}\epsilon_{ki3}+\delta_{j3}\delta_{k3}\epsilon_{li3} + \delta_{i3}\delta_{l3}\epsilon_{kj3} + \delta_{i3}\delta_{k3}\epsilon_{lj3})
\label{S0.3}
\end{aligned}
\end{equation}
In Eq.~(\ref{S0.3}) Einstein summation is not used.  Note that Eq.~(\ref{S0.3}) is not the only possible viscosity tensor that gives Eq.~(\ref{S0.2}). In our DNS we assume $\eta^o_1 = -2\eta^o_2$ and $\eta^o_2 = \rho_0 \nu_{\odd}$. In this case the force due to odd viscosity reduces to $\rho_0\nu_{\rm odd}[\hat{\bm e}_z \times \Delta \bm u + \nabla(\partial_y u_x-\partial_x u_y)]$, where the gradient term can be absorbed into the pressure. Equation~(\ref{S0.2}) then reduces to
\begin{equation}
\begin{aligned}
D_t \bm u = -\nabla P + \nu \Delta \bm u + \nu_{\rm odd}\hat{\bm e}_z\times \Delta \bm u + \bm f
\label{S0.4}
\end{aligned}
\end{equation}
which is Eq.~(1) of the main text. 

In Ref.~\cite{khain2022stokes}, it is proved that the asymmetric viscosity tensor $\eta^A_{ijkl}$ does not generate or dissipate energy. For an intuitive understanding, we consider the following reduced dynamical equation of the odd viscosity term:
\begin{equation}
\begin{aligned}
\partial_t \bm u = \nu_{\rm odd}\hat{\bm e}_z\times \Delta \bm u 
\label{S0.5}
\end{aligned}
\end{equation}
We perform Fourier transform $\bm u = \sum_{\bm k} \tilde {\bm u}(\bm k,t) e^{i\bm k \cdot \bm x}$, which leads to 
\begin{equation}
\begin{aligned}
\partial_t \begin{bmatrix}
        \tilde {u}_x(\bm k,t)\\
        \\
        \tilde {u}_y(\bm k,t)\\
        \\
        \tilde {u}_z(\bm k,t)
    \end{bmatrix}  = \nu_{\rm odd} k^2\begin{bmatrix}
        0 & -1 & 0\\
        \\
        1 & 0 & 0\\
        \\
        0 & 0 & 0
    \end{bmatrix} \begin{bmatrix}
        \tilde {u}_x(\bm k,t)\\
        \\
        \tilde {u}_y(\bm k,t)\\
        \\
        \tilde {u}_z(\bm k,t)
    \end{bmatrix}
\label{S0.6}
\end{aligned}
\end{equation}
The antisymmetric, non-Hermitian dynamical matrix has imaginary eigenvalues $\pm i\nu_{\rm odd} k^2$, which is distinct from that of positive viscosity (negative eigenvalues) and negative viscosity (positive eigenvalues). Equation~(\ref{S0.6}) has oscillatory solutions $\tilde{\bm u}(\bm k,t) \sim e^{\pm i\nu_{\rm odd}k^2 t}$, i.e., energy is neither generated nor  dissipated by odd viscosity. Note that in Eq.~(\ref{S0.5}) the incompressibility $\nabla \cdot \bm u =0$ is not included. Imposing the incompressibility changes the eigenvalues to $\pm i\nu_{\rm odd} |\bm k| k_z$~\cite{khain2022stokes,de2024pattern}. 

Above we have shown that odd viscosity is non-dissipative. As a linear viscosity term, it does not directly affect the nonlinear triadic interaction. Hence, the energy balance equation follows that of the ordinary turbulence~\cite{davidson2015turbulence,Alexakis2018,verma2019energy}:
\begin{equation}
    \label{Ek_dynamics_meth}
    \partial_t E = - 2 \nu k^2 E - T + F, 
\end{equation}
in which
\begin{equation}
    \label{nl_transfer}
    T(\bm{k},t)=\textrm{Im} \sum_{\bm{p}+\bm{q}=\bm{k}} \tilde{u}_i^*(\bm{k},t) P_{ij}(\bm{k})q_\ell \tilde{u}_\ell(\bm{p},t)\tilde{u}_j(\bm{q},t).
\end{equation}
This term describes the non-linear energy transfer between scales, while $F = \vec{u}^* \cdot \vec{f}$ corresponds to energy injection by the forcing term $\vec{f}$. 
The term $-2 \nu k^2 E$ represents standard viscous dissipation.
In Eq.~\eqref{nl_transfer}, the sum runs on momenta $\vec{p}$ and $\vec{q}$ such that $\vec{p}+\vec{q} = \vec{k}$, and $P_{i j}(\vec{k}) = \delta_{ij} - k_i k_j/k^2$ is the projector on incompressible flows \cite{Dar2001,Alexakis2005,Mininni2005,Alexakis2018,verma2019energy}.

\section{Simulation Details}
\subsection{Navier-Stokes}
We perform direct numerical simulations (DNS) of the Navier-Stokes equation incorporating odd viscosity (Eq.~(1) in the main text) within a cubic domain of size $L=2\pi$, employing periodic boundary conditions. Our approach utilizes a pseudo-spectral method coupled with Adams-Bashforth time-stepping and a 2/3-dealiasing rule \cite{Peyret2002}. The dissipative and odd viscous terms are integrated exactly using integrating factors. The forcing $\vec{f}(t,\vec{k})$ acts within a wavenumber band $k\in [k_\text{in},k_\text{in}+1]$, with random phases delta-correlated in space and time, ensuring a consistent average energy injection rate $\epsilon = \langle \vec{u} \cdot \vec{f} \rangle$. This forcing possesses a zero mean component $\langle \vec{f}(t,\vec{k}) \rangle = \vec{0}$ and covariance $\langle \vec{f}(t,\vec{k}) \cdot \vec{f}(t',\vec{k}') \rangle = \epsilon \delta(t-t') \delta(\vec{k}-\vec{k}')$. The time-step is selected to resolve the fastest odd wave with frequency $\tau_{\odd,\text{max}}^{-1} = \nu_\odd k_{\text{max}}^2$, where $k_{\text{max}}$ denotes the highest resolved wavenumber in the domain. We observe that stable integration requires a time step $\Delta t \lesssim 0.1\tau_{\odd,\text{max}}$.

Since the forcing is helicity-free, and odd viscosity conserves helicity, the net total helicity of the flow remains identically zero.

The parameters used for the simulations in the main text are as in Ref.~\cite{de2024pattern}. In Section~\ref{sec3C} we show additional simulations with increased resolution and hyperdissipation to further extend the inertial range. All simulation parameters are provided in Tab.~\ref{tab:input}.

\begin{table*}[h!]
\caption{\label{tab:input}Parameters that are used for DNS in the main text (I) and in Section~\ref{sec3C} (II). Listed are the average energy injection rate $\langle \epsilon \rangle$, normal shear viscosity $\nu$, odd viscosity $\nu_\odd$, injection wavenumber $k_\text{in}$, odd viscosity wavenumber $k_\odd=\epsilon^{1/4}\nu_\odd^{-3/4}$, grid resolution $N^3$, Kolmogorov length $\ell_\nu=\epsilon^{-1/4}\nu^{3/4}$, total simulation time $T$, time-step $\Delta t$, Kolmogorov time $\tau_\nu=\epsilon^{-1/2}\nu^{1/2}$ and the \textit{a posteriori} integral scale Reynolds number without odd viscosity $\text{Re}$.}
\begin{tabular}{cccccccccccc}
\toprule
&$\langle\epsilon\rangle$&$\nu$&$\nu_\odd$&$k_\text{in}$&$k_\odd$&$N^3$&$\ell_\nu/(L/N)$&$T$&$\Delta t$&$\tau_\nu/\Delta t$&$\text{Re}$\\
\midrule
I \rule{0pt}{10pt} & $1.4\times10^{-5}$ & $9.4\times10^{-6}$ & $[0.3 - 2.4]\times10^{-3}$ & $3$ & $[27-6]$ & $768^3$ & $0.34$ & $2\times10^3$ & $[1.0-0.2]\times10^{-2}$ & $[82-410]$ & $1.3\times10^3$\\
II & $1.0$ & $[2-20]\times10^{-10}$ (*) & $2.0\times10^{-2}$ & $2$ & $18$ & $1536^3$ & $0.35$ & $0.6$ & $5\times10^{-5}$ & $200$ & $2.5\times10^3$ \\
\bottomrule
\multicolumn{12}{l}{(*) hyperdissipation (quadratic Laplacian)}
\end{tabular}
\end{table*}

\subsection{Shell Model}
\begin{figure*}[t]
\centering
\hspace{-2em}\includegraphics[width = 0.6\columnwidth]{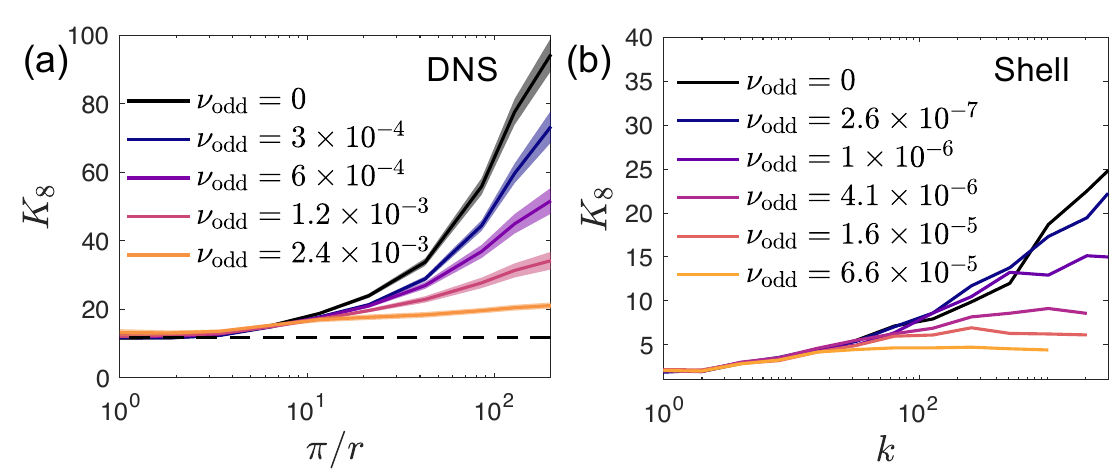}
	\caption{$K_8=S_8/S_4^2$ in DNS (a) and for the shell model (b).  
	}
	\label{fig.S7}
\end{figure*}

In the two-channel helical shell model, shells are logarithmically equispaced with $k_n = k_0 \lambda^n$. The dynamic equation reads (Eq. (3) of the main text)
\begin{equation}
\begin{aligned}
\partial_t u_{n}^{\pm}= B_{n}^{\pm,\rarrow} + B_{n}^{\pm,\larrow}+f_{n}^{\pm}-\nu k_n^2 u_{n}^{\pm},
\label{S1.1}
\end{aligned}
\end{equation}
where $f_n$ is the forcing. 
\begin{equation}
\begin{aligned}
B^{\pm,\rarrow}_n&={a}^{\pm,\rarrow}_n u^{\pm}_{n+2}u_{n+1}^{\mp*}+ b^{\pm,\rarrow}_nu^{\mp}_{n+1}u_{n-1}^{\mp*}- c^{\pm,\rarrow}_nu^{\mp}_{n-1}u_{n-2}^{\pm} \,,\\
B^{\pm,\larrow}_n&= a^{\pm,\larrow}_n u^{\pm}_{n+2}u_{n+1}^{\pm*}+ b^{\pm,\larrow}_nu^{\pm}_{n+1}u_{n-1}^{\pm*}-  c^{\pm,\larrow}_nu^{\pm}_{n-1}u_{n-2}^{\pm}
\label{S1.2}
\end{aligned}
\end{equation}
are the nonlinear transfer of heterochiral and homochiral channels.
\begin{equation}
\begin{aligned}
{a}^{\pm,\rarrow}_n &=k_n g_n^{\pm,\rarrow}=k_n\left(1+\frac{\nu_{\rm odd}k_n}{|u_{n+2}^{\pm}|+|u_{n+1}^{\mp}|+|u_{n}^{\pm}|}\right)^{-1}\,, \\
{a}^{\pm,\larrow}_n &=\beta k_n g_n^{\pm,\larrow}=\beta k_n\left(1+\frac{\gamma\nu_{\rm odd}k_n}{|u_{n+2}^{\pm}|+|u_{n+1}^{\pm}|+|u_{n}^{\pm}|}\right)^{-1}
\label{S1.3}
\end{aligned}
\end{equation}
are the wave-affected transfer coefficients. 
$b_n$ and $c_n$ are determined from the conservation laws of the energy $E=\sum_n|u_n^+|^2+\sum_n|u_n^-|^2$ and the helicity $H=\sum_nk_n|u_n^+|^2-\sum_nk_n|u_n^-|^2$~\cite{benzi1996helical}, which leads to
\begin{equation}
\begin{aligned}
b^{\pm,\rarrow}_n &= -(1-\frac 1 \lambda ) a^{\mp,\rarrow}_{n-1}\,,\\
c^{\pm,\rarrow}_n &= -\frac 1 \lambda a^{\pm,\rarrow}_{n-2}\,,\\
b^{\pm,\larrow}_n &= -(1+\frac 1 \lambda )a^{\pm,\larrow}_{n-1}\,,\\
c^{\pm,\larrow}_n &= \frac 1 \lambda  a^{\pm,\larrow}_{n-2}\,.
\label{S1.4}
\end{aligned}
\end{equation}
The energy balance equation reads~\cite{Biferale2003,verma2019energy}
\begin{equation}
\begin{aligned}
\partial_t E_n=\Pi_{n-1}-\Pi_n -\nu k_n^2 E_n
\label{S1.8}
\end{aligned}
\end{equation}
$\Pi_n$ is the total energy flux
\begin{equation}
\begin{aligned}
\Pi_n =\Pi_n^{\rarrow} + \Pi_n^{\larrow}\,,
\label{S1.7}
\end{aligned}
\end{equation}
where
\begin{equation}
\begin{aligned}
\Pi_n^{\rarrow}&={\rm Im}\left({a}^{+,\rarrow}_n u^{+*}_{n+2}u_{n+1}^{-}u^{+}_{n}+{a}^{-,\rarrow}_n u^{-*}_{n+2}u_{n+1}^{+}u^{-}_{n}-c_{n+1}^{-,\rarrow}u^{+*}_{n+1}u_{n}^{-}u^{+}_{n-1}
-c_{n+1}^{+,\rarrow}u^{-*}_{n+1}u_{n}^{+}u^{-}_{n-1}\right)\,,\\
\Pi_n^{\larrow}&={\rm Im}\left({a}^{+,\larrow}_n u^{+*}_{n+2}u_{n+1}^{+}u^{+}_{n}+{a}^{-,\larrow}_n u^{-*}_{n+2}u_{n+1}^{-}u^{-}_{n}-c_{n+1}^{-,\larrow}u^{-*}_{n+1}u_{n}^{-}u^{-}_{n-1}
-c_{n+1}^{+,\larrow}u^{+*}_{n+1}u_{n}^{+}u^{+}_{n-1}\right)
\label{S1.5}
\end{aligned}
\end{equation}
are the heterochiral and homochiral components of the flux, respectively. 
In the main text we use a velocity-based structure function $S_p(k_n)=\langle (|u_{n-1}^+| |u_{n}^+| |u_{n+1}^{+}|)^{p/3}\rangle$. The definition reduces the 3-shell oscillation that is observed in shell models~\cite{Lvov1998ImprovedSM}. A flux-based structure function $S^\Pi_p(k_n)=\langle |\Pi_n/k_n|^{p/3}\rangle$ is also used in Sec.~\ref{sec4}~\cite{Lvov1998ImprovedSM,de2024extreme}. In the main text we mainly focus on the kurtosis $K=S_4/S_2^2$. We also explore higher order `kurtosis' $K_8=S_8/S_4^2$, which shows that intermittency is suppressed for structure functions up to $p=8$ in both DNS and the shell model (Fig.~\ref{fig.S7}). 
\begin{figure*}[b]
\centering
\hspace{-2em}\includegraphics[width = 1\columnwidth]{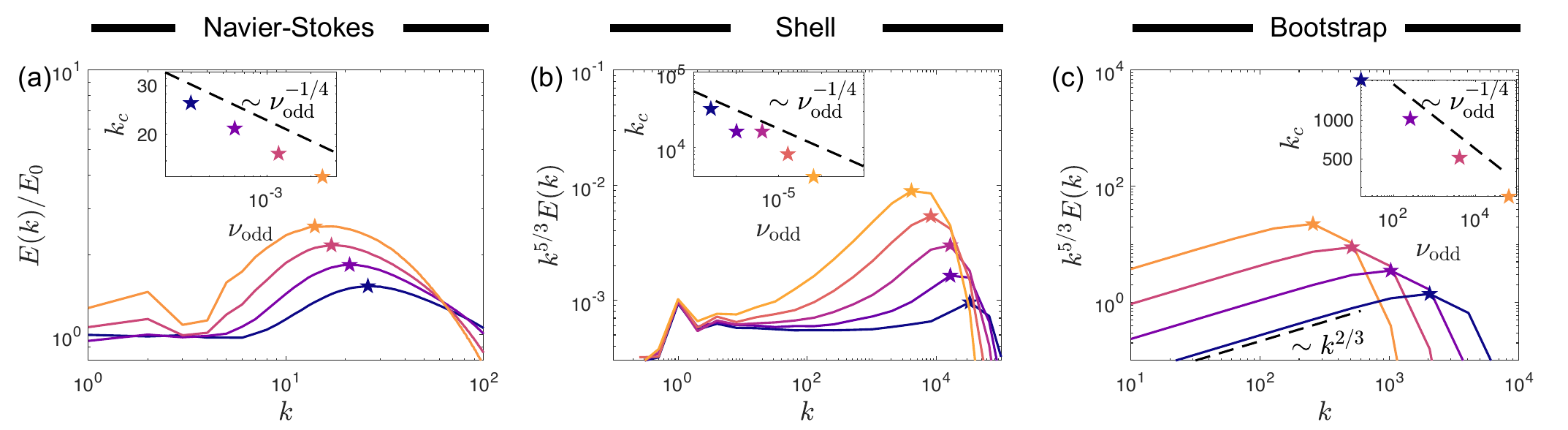}
	\caption{Wavelength selection in odd turbulence. (a) Rescaled energy spectrum $E/E_0$ ($E_0$ is the spectrum for $\nu_{\rm odd}=0$) vs wavenumber in DNS for $\nu_{\rm odd}=3\times 10^{-4},6\times 10^{-4},1.2\times 10^{-3},  2.4\times 10^{-3}$. (b) Rescaled energy spectrum vs wavenumber in the shell model for $\nu_{\rm odd}= 2.6\times 10^{-7}, 1.0\times 10^{-6}, 4.1\times 10^{-6}, 1.6\times 10^{-5}, 6.6\times 10^{-5}$.  (c) Rescaled energy spectrum estimated from the bootstrap method for $\nu_{\rm odd}=16, 256, 4096, 65536$, $\nu=1$ and $\epsilon = 1$. In  (a-c), stars indicate condensation scale $k_c\sim \nu_{\rm odd}^{-1/4}$. 
	}
	\label{fig.S2}
\end{figure*}

We conduct numerical simulations of the two-channel helical shell models w using the 4th order exponential-Runge-Kutta scheme. See Tab.~\ref{tab:input_shell} for parameters.

\begin{table*}[h!]
\caption{\label{tab:input_shell}Parameters that are used for the helical shell model in the main text (I) and in Section~\ref{sec3D} (II). Listed are the energy injection rate $\epsilon$, normal shear viscosity $\nu$, odd viscosity $\nu_\odd$, homochiral parameters $\beta$ and $\gamma$, wave number unit $k_0$, odd viscosity wavenumber $k_\odd=\epsilon^{1/4}\nu_\odd^{-3/4}$, total shell number $N$, total simulation time $T$, time-step $\Delta t$. Note that parameters (II) of the shell model mimic parameters (I) of DNS.}
\begin{tabular}{cccccccccccc}
\toprule
&$\epsilon$&$\nu$  &$\nu_\odd$ & $\beta$ & $\gamma$ &$\lambda$&$k_0$&$N$&$T$&$\Delta t$ \\
\midrule
I \rule{0pt}{10pt} & $1\times10^{-5}$ & $1\times10^{-9}$   & $[0.26-66]\times10^{-6}$ & $0.3$ & $0.2$ & $2$ & $1/8$ & $30$ & $2.5\times 10^5$ & $5\times 10^{-4}$ \\
II  & $1.4\times10^{-5}$ & $9.4\times10^{-6}$   & $[0.3 - 2.4]\times10^{-3}$ & $0.3$ & $0.2$ & $1.618$ & $0.708$ & $30$ & $2.5\times 10^5$ & $5\times 10^{-3}$\\ 
\bottomrule
\end{tabular}
\end{table*}

\begin{figure*}[t]
\centering
\hspace{-2em}\includegraphics[width = 1\columnwidth]{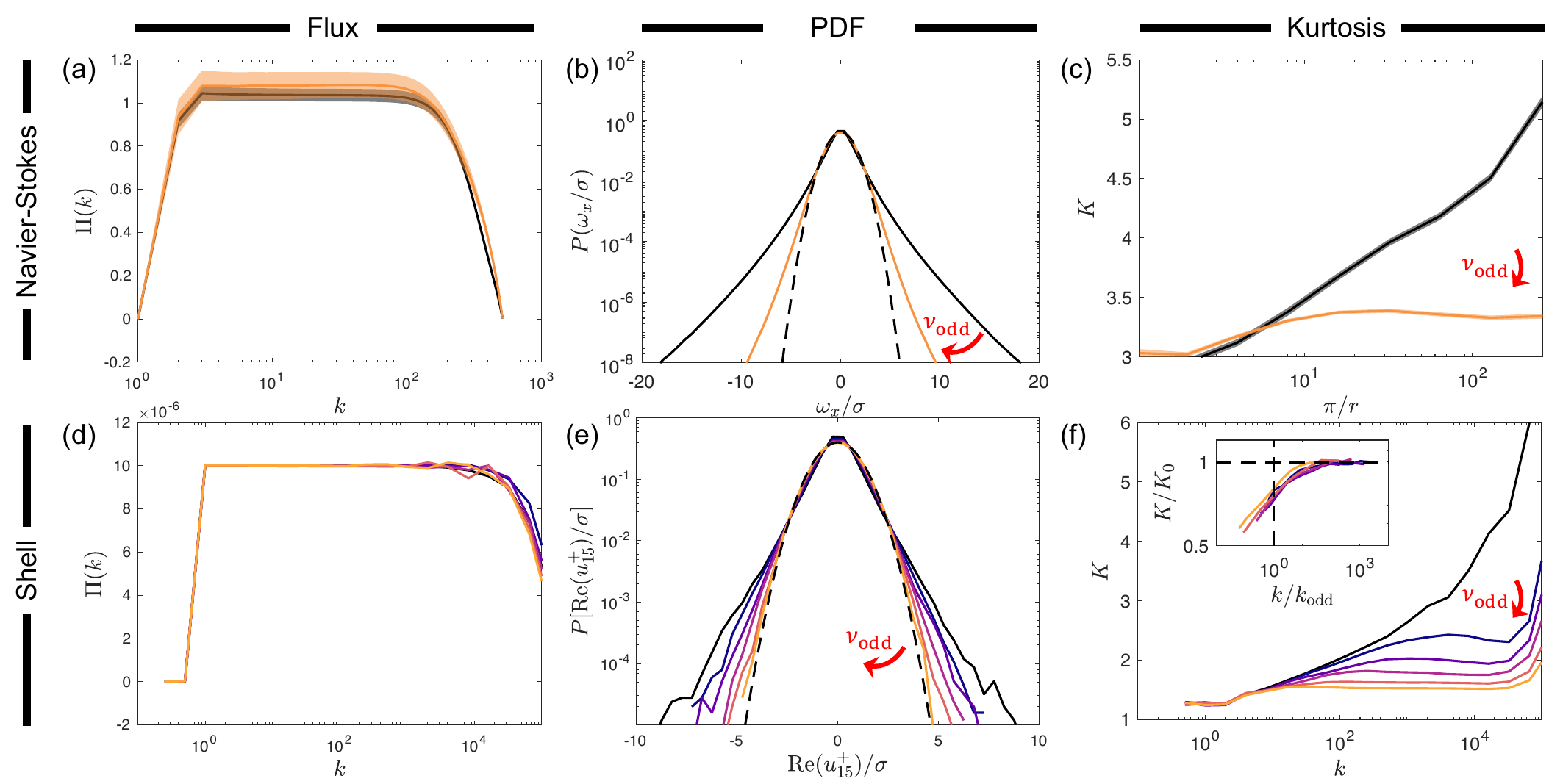}
	\caption{Simulation results with extended inertial range. The value of $\nu$ is reduced when odd viscosity is present, such that the dissipation scale is unchanged. (a) Energy flux in DNS for $\nu_{\rm odd}=0,  2.0\times 10^{-2}$. (b) Probability distribution of $x$-direction vorticity in DNS (renormalized by the standard deviation), for same $\nu_{\rm odd}$ values as in (a). Gaussian distribution is shown in dashed line for comparison. (c) Kurtosis $K$ in DNS, for same $\nu_{\rm odd}$ as in (a). (d) Energy flux in the shell model for $\nu_{\rm odd}=0,  2.6\times 10^{-7}, 1.0\times 10^{-6}, 4.1\times 10^{-6}, 1.6\times 10^{-5}, 6.6\times 10^{-5}$. (b) Probability distribution of ${\rm Re}(u_{15}^+)$ in the shell model (renormalized by the standard deviation), for same $\nu_{\rm odd}$ values as in (d). Gaussian distribution is shown in dashed line for comparison. (c) Kurtosis $K$ in the shell model, for same $\nu_{\rm odd}$ as in (d). 
	}
	\label{fig.S1}
\end{figure*}

\subsection{Dissipation Scale and Extended Inertial Range}
\label{sec3C}
As discussed in Ref.~\cite{de2024pattern} and the main text, odd viscosity affects the dissipation scale $k_c$ with $k_c\sim \epsilon^{1/4}\nu^{-1/2}\nu_{\rm odd}^{-1/4}$. This change of the dissipation scale is due to the change of the energy spectrum caused by odd viscosity: odd viscosity arrest direct cascade and changes the spectrum from $E(k)\sim k^{-5/3}$ to $E(k)\sim k^{-1}$, which enhances the dissipation rate at given $k$ simultaneously. Therefore, when odd viscosity is present the dissipation dominates at larger lengthscales $k_c$. In Fig.~\ref{fig.S2}(a,b) we plot the rescaled energy spectrum $E(k)/E_0$ in both DNS and the shell model ($E_0$ is the spectrum for $\nu_{\rm odd}=0$). We find that $E(k)/E_0$ increases with $k$ for $k_\odd <k<k_c$ because of the arrested cascade.  For $k>k_c$, $E(k)/E_0$ decreases with $k$ due to the dissipation. This effectively leads to a wavelength selection phenomenon, which is also known as pattern formation~\cite{de2024pattern}. 

Here we provide a simple bootstrap method which reproduces the pattern formation in odd turbulence. Similar to the shell model, we divide the system into logarithmically distributed shells $k_n=k_0 2^n$. The velocity fluctuations in each shell is $u_n$ with shell energy $E_n= u_{n}^2$. Assuming $k_n>k_\odd$, the wave-affected energy transfer rate from shell $n$ to $n+1$ is $\epsilon_n\sim \nu_{\rm odd}^{-1} u_n^4=\nu_{\rm odd}^{-1} E_n^2$ (see Sec.~\ref{sec3}). In the steady state, the 
energy conservation at shell $n+1$ requires $\epsilon_n=\epsilon_{n+1}+\nu k_{n+1}^2 E_{n+1}$, i.e., the energy injected in the shell equals to the energy ejected from the shell plus dissipation. Because this equation holds for all shells in the steady state, an estimated energy spectrum can be calculated from an iteration of the energy conservation of all shells, assuming $\epsilon_0=\epsilon$. The results are shown in Fig.~\ref{fig.S2}c. We indeed find pattern formation at $k_c$, and the dependence of $k_c$ on $\nu_{\rm odd}$ is consistent with our expectation. 

Because $k_c$ decreases for increasing $\nu_{\rm odd}$, the inertial range also shrinks, and the statistics of the kurtosis is affected more by the dissipation when odd viscosity is present. To avoid this effect, we conduct simulations with extended inertial range for both DNS and the shell model. This is achieved by reducing the viscosity to $\tilde\nu\sim\nu^{3/2}\nu_{\rm odd}^{-1/2}$ for large $\nu_{\rm odd}$. This approach ensures that the inertial range remains approximately same for different $\nu_{\rm odd}$ values, see Fig.~\ref{fig.S1}(a,d). The suppression of intermittency is again observed in both DNS and the shell model (Fig.~\ref{fig.S1}(b,c,e,f)). Interestingly, in DNS we observe a weak non-monoticity of the kurtosis, in alignment with the shell model result and the prediction of the modified Parisi-Frisch theory.

\subsection{Results for smaller scale separation}
\label{sec3D}
In the main text, for both the modified Paris-Frisch theory and the shell model we use much larger scale separation than what is accessible in DNS. Such a large scale separation allows us to separate the results into clearly-defined regimes: for $k_{\rm in}\ll k\ll k_{\rm odd}$, there is a direct cascade of ordinary turbulence; for $k_{\rm odd}\ll k\ll k_{c}$, there is a direct cascade of odd-wave-affected turbulence. Such a distinction between the two regimes requires a sufficiently large separation between the injection scale $k_{\rm in}$, the cross-over scale $k_{\rm odd}$ and the dissipation scale $k_c$, and the scale separation in DNS is limited by computational expense. 

In this section we provide results for the modified Paris-Frisch theory and the shell model, using similar scale separation as that in DNS. The results exhibit similar qualitative features as that observed in DNS, validating the two theoretical approaches. 

\begin{figure*}[h]
\centering
\hspace{-2em}\includegraphics[width = 1\columnwidth]{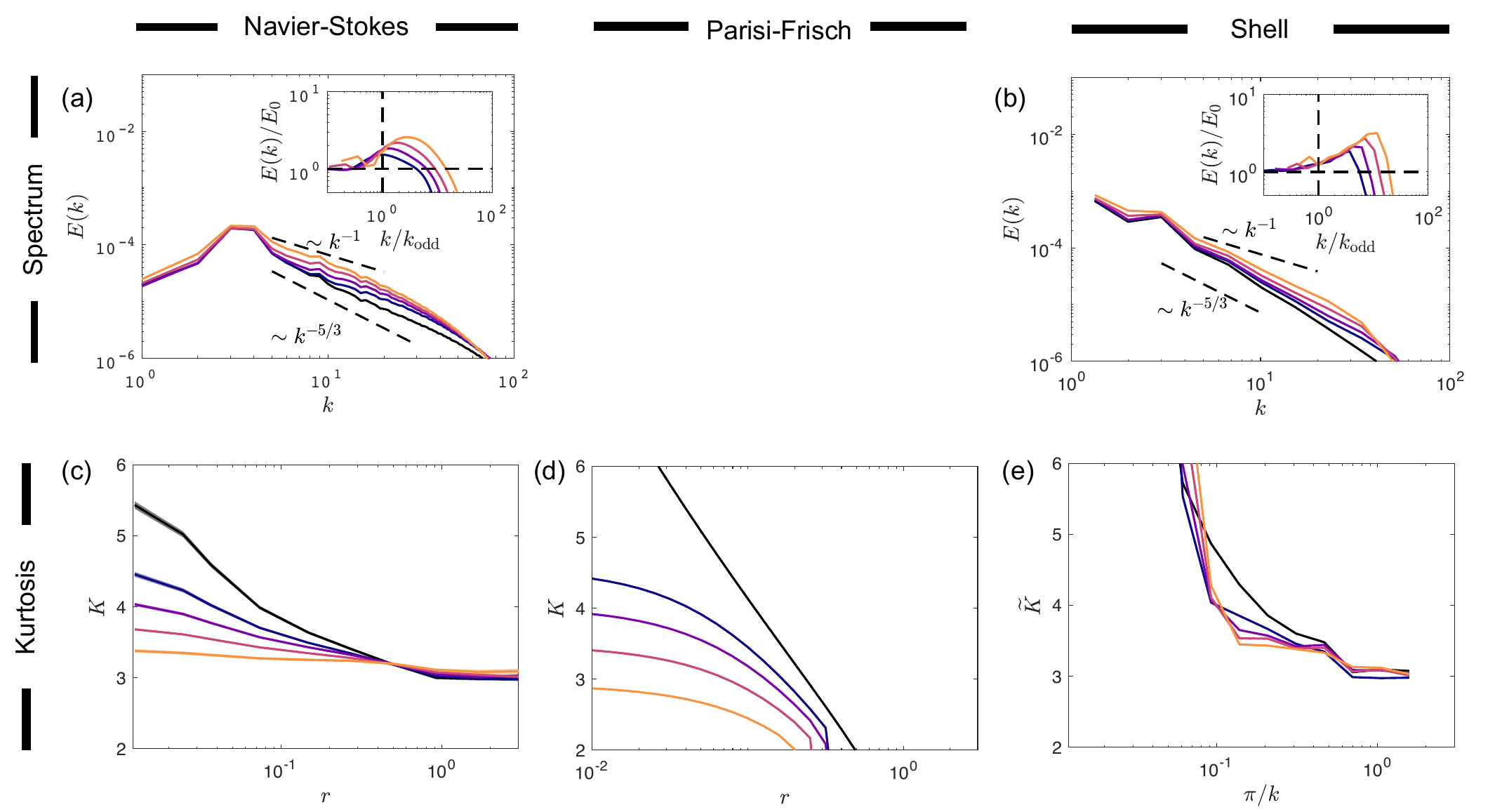}
	\caption{Comparison between DNS, modified Parisi-Frisch and shell model results, for same scale separation as that in DNS. (a) and (c) Energy spectrum and kurtosis in DNS for $\nu_{\rm odd}=0, 3\times 10^{-4}, 6\times 10^{-4}, 1.2\times 10^{-3}, 2.4\times 10^{-3}$, same as that in Fig.~3 of the main text. (d) Kurtosis in the modified Parisi-Frisch theory for $\nu_{\rm odd}=0, 1.2\times 10^{-2}, 2.5\times 10^{-2}, 4.9\times 10^{-2}, 9.9\times 10^{-2}$. (b) and (d) Energy spectrum and rescaled kurtosis in the shell model for $\nu_{\rm odd}=0, 3\times 10^{-4}, 6\times 10^{-4}, 1.2\times 10^{-3}, 2.4\times 10^{-3}$. In (e) kurtosis is rescaled with $\widetilde K = 2.4 K$, which ensures that $\widetilde K \approx 3$ at the forcing scale. 
	}
	\label{fig.S8}
\end{figure*}

Specifically, for the modified Parisi-Frisch theory we ensure the same odd viscosity scale $k_{\odd}\in[27-6]$ as that in DNS. In the modified Parisi-Frisch theory, we define $k_{\odd}=1/r_{\odd}(h=1/3)$, where $r_{\odd}(h)=\nu_{\odd}^{1/(1+h)}$. This corresponds to  $\nu_{\odd}\in [1.2-9.9]\times 10^{-2}$. As shwon in Fig.~\ref{fig.S8}(d), the modified Parisi-Frisch theory reproduces the suppression of kurtosis by odd viscosity. Note that for $r$ close to $1$ we observe an unphysical discontinuity. This is because the original  Parisi-Frisch theory is designed only for the asymptotic behavior of the intermittency at small scales, and is ill-defined at large scales. 

For the shell model we use exactly the same $\epsilon$, $\nu$ and $\nu_{\odd}$ values as that used in DNS. The energy spectrum of the shell model reproduces the arrest of direct cascade, see Fig.~\ref{fig.S8}(b). The suppression of kurtosis by odd viscosity is also observed in Fig.~\ref{fig.S8}(e) for scale above the dissipation scale. For scale below the dissipation scale, we find anomalous increase of kurtosis, which is a known issue for most shell models~\cite{benzi2009fully}.

\section{Intermittency suppressed by Breakdown of multiple scale invariances}
\label{sec3}
As mentioned in the main text, intermittency is suppressed in odd-wave-affected turbulence, because the multiple scale invariances of the Navier-Stokes equation are broken by odd viscosity. In this section we detail how the multiple scale invariances are broken by the parity-breaking waves produced by odd viscosity.

We start with the inertial-range Navier-Stokes equation
\begin{equation}
\begin{aligned}
\partial_t \bm u + \bm u \cdot\nabla\bm u= -\nabla P  + \nu_{\rm odd}\hat{\bm e}_z\times \Delta \bm u \,,
\label{S4.1}
\end{aligned}
\end{equation}

In the absence of $\nu_{\rm odd}$, the inertial Navier-Stokes equation is invariant under global scale transformation~\cite{ParisiFrisch1985}
\begin{equation}
\begin{aligned}
\bm x\to \lambda \bm x, \bm u \to\lambda^h \bm u,  t\to \lambda^{1-h}t\,,
\label{S4.2}
\end{aligned}
\end{equation}
for arbitrary exponent $h$. We define this property as the multiple {\it global} scale invariances, which implies that singularities (flow with local self-similar structure) $\delta_r u\sim r^h$ with arbitrary $h$ can develop in turbulent flow. In Parisi-Frisch multifractal formalism, the turbulent flow contains singularities with a range of exponents $h$ that are distributed in different locations in the 3D space, which further leads to intermittency~\cite{ParisiFrisch1985,Frisch1995}. 

For non-zero $\nu_{\rm odd}$, because of the odd term, the {\it global} scale invariance only holds for $h=-1$.  All other {\it global} scale invariances are broken by the linear odd term. This conclusion can be generalized to any wave-turbulence system described in Ref.~\cite{Zakharov1992Kolmogorov}, where a wave-generating linear term breaks multiple {\it global} scale invariances. However, the breaking of multiple {\it global} scale invariances does not necessarily suppress intermittency. As an example, we consider a flow described by Eq.~(\ref{S4.1}) and artificially block the energy transfer between all modes with $k_z \neq 0$. This enforces a 2D flow with $k_z=0$ on which odd viscosity has no effect at all, and 2D singularities with arbitrary $h$ are still allowed by scale symmetry (although in reality 2D turbulence is non-intermittent for a different reason~\cite{paladin1987anomalous,eyink2006onsager}, see discussion at the end of the section). In other words, while the {\it global} scale invariance of Eq.~(\ref{S4.1}) only holds for $h=-1$, its 2D part has  {\it partial} scale invariances for all $h$. 

This example shows that what matters is not the {\it global} scale invariance of the complete Navier-Stokes equation, but is the {\it partial} scale invariance of the energy-transferring part of the equation. The non-energy-transferring part of the Navier-Stokes equation does not contribute to the turbulent cascade, hence it does not affect the intermittency. 

To identify the energy-transferring part of Eq.~(\ref{S4.1}), we perform helical decomposition,
\begin{equation}
\begin{aligned}
\bm u(t,\bm x) &= \sum_{\bm k}\sum_{s=\pm}u_s(t,\bm k)\bm j^{s}(\bm k)e^{i\omega_s(\bm k) t + i\bm k\cdot\bm x}\,,
\label{S4.4}
\end{aligned}
\end{equation}
where $\bm j^{\pm}(\bm k)=\hat{\bm e}(\bm k)\times (\bm k/|\bm k|)\pm i\hat{\bm e}(\bm k) $ with $\hat{\bm e}(\bm k)=\hat{\bm e}_z\times \bm k/|\hat{\bm e}_z\times \bm k|$ and frequency 
\begin{equation}
\begin{aligned}
\omega_{\pm}(\bm k)=\pm \nu_{\rm odd}k_z|\bm k|\,.
\label{S4.5}
\end{aligned}
\end{equation}
$\pm$ refers to the definite helicity of each helical mode. The explicit dependence of the frequency on the helicity is a result of the broken parity symmetry in odd-viscous fluids.  We rewrite Eq.~(\ref{S4.1}) (in the inertial range) as
\begin{equation}
\begin{aligned}
\partial_t u_{s_k}=\sum_{\substack{\bm k + \bm p + \bm q =0\\s_p, s_q = \pm}}C_{k|p,q}e^{i\bar\omega(\bm k,\bm p,\bm q) t}u_{s_p}^*u_{s_q}^*\,.
\label{S4.6}
\end{aligned}
\end{equation}
Here, a triad with wave vectors $(\bm k,\bm p, \bm q)$ and definite helicities $(s_k,s_p,s_q)$ is a basic unit for energy transfer. 
\begin{equation}
\begin{aligned}
C_{k|p,q}=-\frac 1 4 (s_p p-s_q q)\left[(\bm j^{s_p}(\bm p) \times \bm j^{s_q}(\bm q))\cdot \bm j^{s_k}(\bm k)\right]^*
\label{S4.7}
\end{aligned}
\end{equation}
is a coefficient that has the dimension of wave number. $\bar\omega(\bm k,\bm p,\bm q)=\omega_{s_k}(\bm k)+\omega_{s_p}(\bm p)+\omega_{s_q}(\bm q)$ is an important frequency of the triad $(\bm k, \bm p, \bm q)$. As we discussed in the main text, energy transfer is effectively blocked when $\bar\omega\tau_{\rm eddy}\gg1$, because the fast oscillation of $e^{i\bar\omega t}$ cancels the average energy transfer. Therefore, we may divide all triads into quasi-resonant triads ($\bar\omega\tau_{\rm eddy}\leq1$) and quasi-non-resonant triads ($\bar\omega\tau_{\rm eddy}>1$). The energy transfer is approximately unaffected by odd viscosity in quasi-resonant triads, and is blocked in quasi-non-resonant triads. 

The energy-transferring part of Eq.~(\ref{S4.1}) is then a decimated Navier-Stokes equation in which only quasi-resonant triads are preserved:
\begin{equation}
\begin{aligned}
\partial_t u_{s_k}=\sum_{\substack{\bm k + \bm p + \bm q =0\\s_p, s_q = \pm\\\bar\omega\tau_{\rm eddy}\leq1}}C_{k|p,q}u_{s_p}^*u_{s_q}^*\,.
\label{S4.8}
\end{aligned}
\end{equation}
While Eq.~(\ref{S4.8}) has a similar form as that of the original Navier-Stokes equation, a different nonlinear dependence on $u$ emerges from the number of resonant triads, because $\tau_{\rm eddy}\sim 1/(ku)$. 

As discussed in the main text, we consider quasi-local triads ($k=|\bm k|\approx|\bm p|\approx |\bm q|$) which dominate the energy cascade. Letting $u_k$ be the average velocity fluctuations at wave number $k$, Eq.~(\ref{S4.8}) is (statistically) simplified to
\begin{equation}
\begin{aligned}
\partial_t u_k\sim g_k ku_k^2\,,
\label{S4.9}
\end{aligned}
\end{equation}
where $g_k$ is the fraction of quasi-resonant triads. For homochiral triads which tansfer energy inversely ($s_p=s_k=s_q$), $\bar{\omega}/\nu_{\rm odd} = s_k k_z|\bm k|+s_p p_z|\bm p| + s_q q_z|\bm q|\approx s_k|\bm k|(k_z+p_z+q_z)=0$ (because $\bm k + \bm q + \bm p =0$), leading to $g_k\approx 1$. Substituting $g_k$ in Eq.~(\ref{S4.9}), we have the following scale invariances of the homochiral channel: 
\begin{equation}
\begin{aligned}
k\to \lambda^{-1} k,  u_k \to\lambda^h u_k,  t\to \lambda^{1-h}t\,,
\label{S4.10}
\end{aligned}
\end{equation}
for arbitrary $h$. This is identical to Eq.~(\ref{S4.2}). 

For heterochiral triads which carries energy forward ($s_p=-s_k=s_q$), $\bar{\omega}/\nu_{\rm odd} = s_k k_z|\bm k|+s_p p_z|\bm p| + s_q q_z|\bm q|\approx 2s_k|\bm k|k_z$.  For given $k$, $\bar{\omega}$ is uniformly distributed in $[-2k^2\nu_{\rm odd},2k^2 \nu_{\rm odd}]$. Thus, $g_k=1$ for $\nu_{\rm odd}\leq u_k/(2k)$ and $g_k=u_k/(2k \nu_{\rm odd})$ for $\nu_{\rm odd}> u_k/(2k)$. In the wave-affected regime, substituting $g_k\sim u_k/k$ in Eq.~(\ref{S4.9}), we have the following scale invariances of the heterochiral channel:
\begin{equation}
\begin{aligned}
k\to \lambda^{-1} k,  u_k \to\lambda^h u_k,  t\to \lambda^{-2h}t\,,
\label{S4.11}
\end{aligned}
\end{equation}
for arbitrary $h$. Note that the scale transformations applied on $t$ are different in Eqs.~(\ref{S4.10},\ref{S4.11}). Because energy is transferred through homochiral and heterochiral channels simultaneously, a singularity must satisfies both Eqs.~(\ref{S4.10},\ref{S4.11}). In that case, the only scaling exponent consistent with both channels is $h=-1$, and the multiple {\it partial} scale invariances are broken. In this case, the flow is statistical self-similar, i.e., $\delta_r u\sim r^h$ only holds for a single $h$. Because the energy transfer is dominated by the forward flux of heterochiral triads, we have $\epsilon\sim g_k ku_k^3\sim \nu_\odd^{-1}u_k^4$. In this case, $u_k\sim\nu_\odd^{1/4} k^0$ and for consistency we have $\delta_r u\sim \nu_\odd^{1/4} r^0$. This conclusion is used in the modified Parisi-Frisch theory in the main text (Eq.~(2)). 

\begin{figure*}[t]
\centering
\hspace{-2em}\includegraphics[width = 0.7\columnwidth]{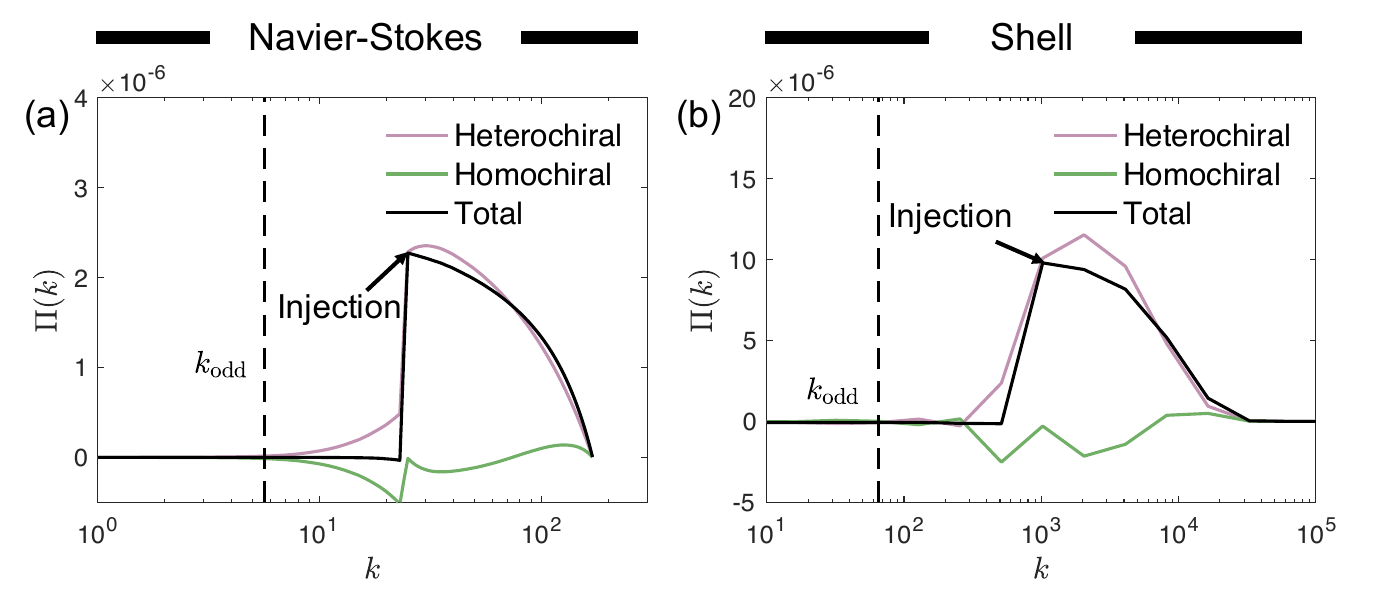}
	\caption{Flux loop for small-scale forcing. (a) Energy flux in DNS for $\nu_{\rm odd}= 2.4\times 10^{-3}$. (b) Energy flux in the shell model for $\nu_{\rm odd}= 6.6\times 10^{-5}$. Forcing is applied on $n=13$. 
	}
	\label{fig.S3}
\end{figure*}

The inconsistent scale invariances of homochiral and heterochiral triads are due to the asymmetric effects of odd viscosity on them, i.e., $\bar{\omega}_{\rm homo}\neq\bar{\omega}_{\rm hetero}$, which originates from the broken-parity symmetry of odd fluids. Indeed, the mechanism discussed above is only valid when the wave frequency explicitly depends on the definite helicity, which requires broken parity. This mechanism can be generalized to any generalized odd waves with frequency $\omega_{\pm}(\bm k)=\pm c_{\rm odd}k_z |\bm k|^{\alpha-1}$, as we discuss in Sec.~\ref{sec2}, or any parity-breaking waves. Their parity-preserving counterparts, e.g., waves with frequency $\omega_{\pm}(\bm k)=c_{r}k_z |\bm k|^{\alpha-1}$ that does not depend on helicity, do not break the multiple {\it partial} scale invariances of the Navier-Stokes equation. This is supported by our shell model results, see Sec.~\ref{sec4}.
Intermittency has also been experimentally reported in parity-preserving wave turbulence, e.g., gravity capilary wave turbulence~\cite{falcon2007observation, lukaschuk2009gravity}.

Another consequence of the asymmetric effect of odd viscosity on homochiral and heterochiral channels is the enhanced flux loop, as shown in the main text. Because the homochiral and heterochiral channels transfer energy in different directions, a weak inverse flux in the homochiral channel exists in the direct cascade of ordinary turbulence~\cite{deusebio2014dimensional}. The inverse flux of the homochiral channel, together with the forward flux of the heterochiral channel, forms a flux loop, which is amplified by odd viscosity for $k>k_{\rm odd}$. More strikingly, when forcing at small scale $k_{\rm in}>k_{\rm odd}$, a flux loop develops at large scale in the absence of a net energy flux~\cite{de2024pattern}. This is also observed in the shell model (Fig.~\ref{fig.S3}).  

Importantly, the breakdown of multiple scale invariances requires zero or small net helicity injection, which ensures that energy is equally distributed on waves with both definite helicities. When large helicity is injected, the flow can be dominated by waves with the same definite helicity as the injected one. In this case, the nonlinear transfer is dominated by the homochiral channel only, and the asymmetry between the two channels has small effect. Indeed, intermittency has been reported in rotating turbulence (parity is broken by the rotation direction) when large helicity is injected~\cite{mininni2010rotating}. 

Another turbulent system that is known to be non-intermittent is 2D turbulence. Although this non-intermittency in 2D has not been fully understood, it must be governed by a different mechanism from what is discussed above. The reason is that the suppression of intermittency in 3D odd turbulence is due to the breakdown of scale invariances. 2D turbulence, on the other hand, has exactly the same scale invariances as that in 3D ordinary turbulence. Hence, intermittency is allowed in 2D from a perspective of scale invariance. It is conjectured that the absence of intermittency in 2D turbulence is related to the energy being transferred from fast degrees of freedom to slower degrees of freedom~\cite{vladimirova2021fibonacci}.

\section{Generalized odd viscosity}
\label{sec2}
In the main text  we discuss the effect of odd viscosity on turbulent fluids, which produces linear waves with frequencies
\begin{equation}
\begin{aligned}
\omega_{\pm}(\bm k)=\pm \nu_{\rm odd}k_z|\bm k|\,.
\end{aligned}
\end{equation}
In this section we generalize the discussion to generalized odd viscosity with wave frequencies 
\begin{equation}
\begin{aligned}
\omega_{\pm}(\bm k)=\pm c_{\rm odd}k_z|\bm k|^{\alpha-1}\,
\label{S3.2}
\end{aligned}
\end{equation}
where $c_{\rm odd}$ is a generalized odd viscosity. $\alpha$ can be any real number. The case for odd viscosity discussed in the main text corresponds to $\alpha=2$. When $\alpha=0$, one can show that Eq.~(\ref{S3.2}) is equivalent to the Navier-Stokes equation for rotating turbulence with rotating frequency $\Omega=c_{\rm odd}/2$~\cite{galtier2003weak}. Importantly, all generalized odd waves are parity-breaking waves, since the frequency explicitly depends on the chirality. 

We can repeat the analysis of the resonance condition for homochiral and heterochiral triads, as we do in the main text and Sec.~\ref{sec3} for the $\alpha=2$ case. For heterochiral triads with locality ($k=|\bm k|\approx|\bm p|\approx |\bm q|$) we find the fraction of resonant triads $g_k=1$ for $c_{\rm odd}\leq u_k/(2k^{\alpha-1})$ and $g_k=u_k/(2k^{\alpha-1} c_{\rm odd})$ for $c_{\rm odd}> u_k/(2k^{\alpha-1})$. This modified the energy transfer rate to $\epsilon\sim c_{\rm odd}^{-1}k^{2-\alpha}u_k^4$, which supports a steady-state, wave-affected solution $u_k\sim \epsilon^{1/4}c_{\rm odd}^{1/4}k^{\alpha/4-1/2}$. The scaling of the wave-affected solution is consistent with that obtained from the weak-turbulence theories~\cite{Zakharov1992Kolmogorov,galtier2000weak,galtier2003weak}. The 'generalized odd' scale $k_{\rm odd}=\epsilon^{1/(3\alpha-2)}c_{\odd}^{3/(2-3\alpha)}$ is found by equating the wave-affected solution and the Kolmogorov solution. The dissipation scale $k_c$ can be estinated from $\epsilon\sim \nu k_c^2 u_{k_c}^2$, which leads to $k_c\sim \epsilon^{1/(\alpha+2)}c_{\rm odd}^{-1/(\alpha+2)}\nu^{-2/(\alpha+2)}$. 

For homochiral triads with locality we again find $g_k=1$, i.e., all local triads are resonant. Here, note that homochiral triads with exact locality do not transfer energy, because $C_{k|p,k}=0$ when $k=p=q$. Therefore, homochiral triads can only transfer energy through weakly non-local triads, which are also weakly affected by the resonance condition when $\alpha \neq 1$. For $\alpha=1$, on the other hand, we find that the resonance condition is satisfied by all homochiral triads regardless of the locality. A similar situation also happens in MHD turbulence, in which the frequencies of the Alf\'ven waves coincide with that of generalized odd waves with $\alpha=1$~\cite{galtier2000weak}. 

For arbitrary $\alpha$ value, the effects of generalized odd viscosity on homochiral and heterochiral channels are asymmetric. Hence, the multiple {\it partial} scale invariances discussed in Sec.~\ref{sec3} are broken for any $\alpha$ in the wave-affected regime.

\subsection{Modified Parisi-Frisch}
We now extend the modified Parisi-Frisch formalism to generalized odd viscosity. In this case, self-similar solutions with $\delta_r u\sim r^h$ at large scales reduce to the wave-affected solution $\delta_r u\sim c_{\rm odd}^{1/4} r^{1/2-\alpha/4}$ (because $u_k\sim \epsilon^{1/4}c_{\rm odd}^{1/4}k^{\alpha/4-1/2}$). The crossover is phenomenologically written as
\begin{equation}
\begin{aligned}
\delta_r u &= r^h\left[1+r_{\rm odd}(h)/r\right]^{h-1/2 + \alpha/4}\,,\\
P_h(r) &= r^{F(h)}\left[1+r_{\rm odd}(h)/r\right]^{F(h)}\,,\\
\label{S2.4}
\end{aligned}
\end{equation}
which is same as Eq.~(2) of the main text. Here, the crossover lengthscale $r_{\rm odd}(h)$ is determined by equating two timescales: the eddy turnover time $\tau=r/\delta_r u$ and the timescale of generalized odd waves $\tau_{\rm odd}=c_{\rm odd}^{-1}r^\alpha$, which gives $r_{\rm odd}(h)=c_{\rm odd}^{1/(h+\alpha-1)}$. We adopt a commonly-used empirical choice of $F(h)$: $F(h)=2-c_1(h-1/9)+c_2(h-1/9)\ln(h-1/9)$, with $c_1=3[[1+\ln(\ln(3/2))]/\ln(3/2)-1]$ and $c_2 = 3/\ln(3/2)$~\cite{she1994universal}. 

The suppression of intermittency at small lengthscales holds for arbitrary $\alpha>2/3$, which exhibits the same qualitative features as the results for $\alpha=2$ shown in the main text. For $\alpha<2/3$, the generalized odd viscosity dominates at large lengthscales, e.g., in rotating turbulence, where Eq.~(\ref{S2.4}) becomes invalid.  

\begin{figure*}[b]
\centering
\hspace{-2em}\includegraphics[width = 1\columnwidth]{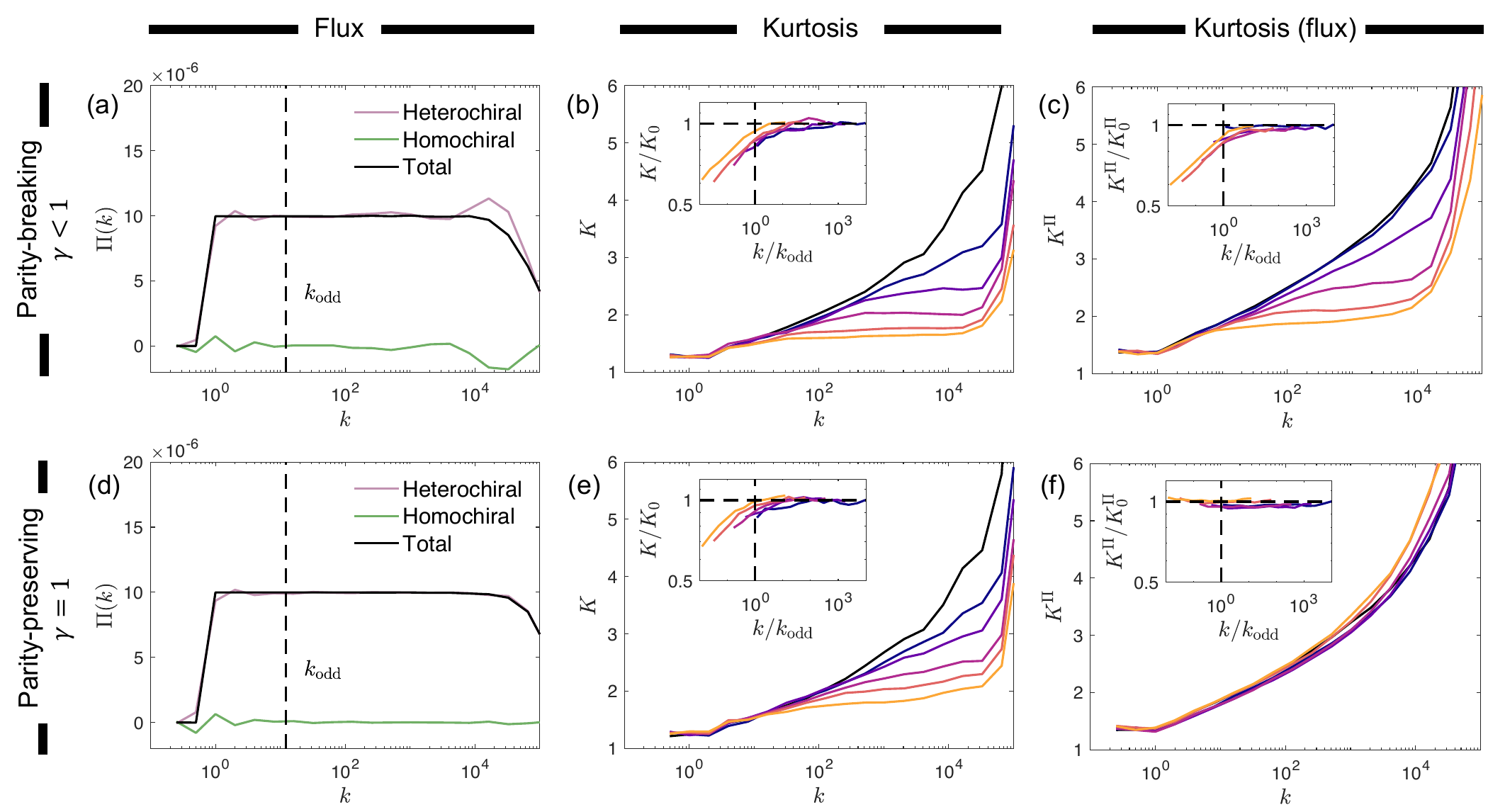}
	\caption{Shell model simulations for wave frequencies $\omega(\bm k)\sim c_{\rm odd} k_z|\bm k|^{\alpha-1}$ with $\alpha=1.5$. Inertial range is extended by reducing the normal viscosity, see Sec.~\ref{sec3C}. (a) Energy flux for parity-breaking waves ($\gamma<1$) and $c_{\rm odd}=8.4\times 10^{-3}$. (b) Velocity-based kurtosis for parity-breaking waves ($\gamma<1$) and $c_{\rm odd}=0,8.3\times 10^{-6},3.3\times 10^{-5},1.3\times 10^{-4},5.2\times 10^{-4},2.1\times 10^{-3},8.4\times 10^{-3}$. (c) Flux-based kurtosis for parity-breaking waves ($\gamma<1$) and same $c_{\rm odd}$ values as in (b). (d) Energy flux for parity-preserving waves ($\gamma=1$) and $c_{\rm odd}=8.4\times 10^{-3}$. (e) Velocity-based kurtosis for parity-preserving waves ($\gamma=1$) and $c_{\rm odd}$ values same as in (b).  (f) Flux-based kurtosis for parity-preserving waves ($\gamma=1$) and same $c_{\rm odd}$ values as in (b). 
	}
	\label{fig.S4}
\end{figure*}

\begin{figure*}[t]
\centering
\hspace{-2em}\includegraphics[width = 1\columnwidth]{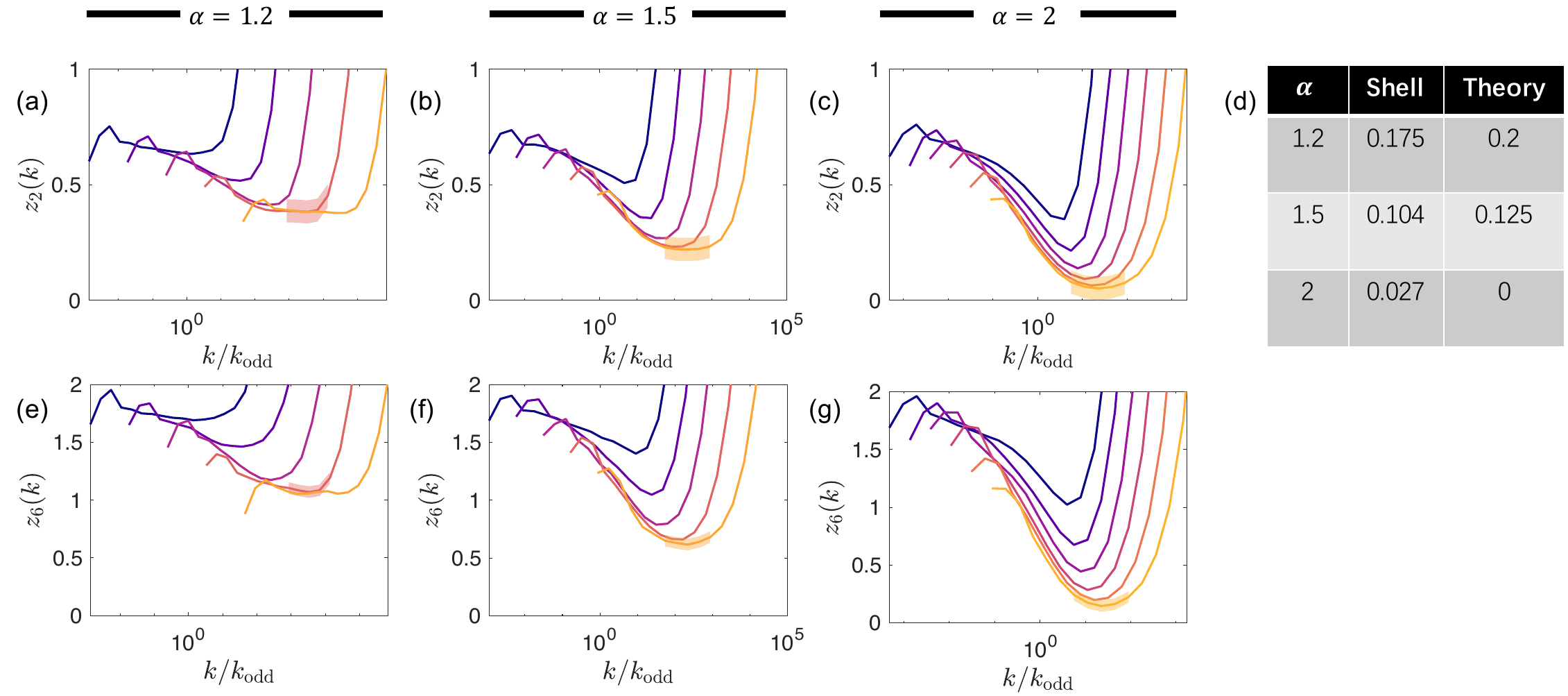}
	\caption{Differential scaling exponent of $S_2$ (a-c) and $S_6$ (e-g) in the shell model with generalized odd viscosity (inducing generalized odd waves with frequency $\omega_{\pm}(\bm k) = \pm c_{\odd} k_z|\bm k|$). Exponents are plotted vs rescaled wavenumber $k/k_\odd$, with $k_{\rm odd}=\epsilon^{1/(3\alpha-2)}c_{\odd}^{3/(2-3\alpha)}$. Results are shown for $c_{\rm odd}/c_0=0, 4^m$, with $c_0 = (\epsilon/2)^{1/2-\alpha/4}\nu^{3\alpha/4-1/2}$.  (a, e) $\alpha=1.2$ and $m=2-6$. (b, f) $\alpha=1.5$ and $m=3-7$. (c, g) $\alpha=2$ and $m=4-9$. (d) Coefficient $h_\alpha=\zeta_p/p$ obtained from the shell model. Theoretical approximation $h_\alpha=1/2-\alpha/4$ is shown for comparison. 
	}
	\label{fig.S6}
\end{figure*}

\subsection{Two-channel helical shell model}
\label{sec2B}
We generalize here the two-channel helical shell model to include generalized odd viscosity. This is achieved by replacing Eq.~(\ref{S1.3}) with
\begin{equation}
\begin{aligned}
{a}^{\pm,\rarrow}_n &=k_n g_n^{\pm,\rarrow}=k_n\left(1+\frac{c_{\rm odd}k_n^{\alpha-1}}{|u_{n+2}^{\pm}|+|u_{n+1}^{\mp}|+|u_{n}^{\pm}|}\right)^{-1}\,, \\
{a}^{\pm,\larrow}_n &=\beta k_n g_n^{\pm,\larrow}=\beta k_n\left(1+\frac{\gamma c_{\rm odd}k_n^{\alpha-1}}{|u_{n+2}^{\pm}|+|u_{n+1}^{\pm}|+|u_{n}^{\pm}|}\right)^{-1}\,.
\label{S2.7}
\end{aligned}
\end{equation}

The value of $\gamma$ controls which channel is affected more by generalized odd viscosity. According to our analysis of resonant triads in the main text, $\gamma<1$ such that odd viscosity has a weaker effect on the homochiral channel. We show here the results for $\alpha=1.5$, which exhibits the same qualitative features as the $\alpha=2$ case in the main text (Fig.~\ref{fig.S4}a,b). 

As shown in Fig.~4(a), generalized odd viscosity modifies the scaling behavior of the structure functions for $k\gg k_{\odd}$. This can be quantitatively described by a differential scaling exponent $z_p(k)=-\diff \log(S_p)/\diff \log(k)$. We plot $z_2$ and $z_6$ in the shell model for different $\alpha$ and $c_{\odd}$ values, vs rescaled wavenumber $k/k_{\odd}$, see Fig.~\ref{fig.S6}a-c, e-g.  We find that $z_p$ changes from the usual multifractal exponents of regular turbulence for $k\ll k_{\odd}$ to a plateau for $k\gg k_{\odd}$. The plateau corresponds to a fully wave-affected regime where $S_p(k_n)\sim k_n^{-\zeta_p}$. We fit the values of $\zeta_p$ using the shaded area, and the results are shown in Fig. 4d of the main text. We then fit $\zeta_p$ using the monofractal formula $\zeta_p=h_\alpha p$. The numerical values of $h_\alpha$ are close to the theoretical approximation $\zeta_p=p(1/2-\alpha/4)$ (Fig.~\ref{fig.S6}d). The theoretical $h_\alpha$ is obtained using the resonant fraction, see below Eq.~(\ref{S3.2}).

\begin{figure*}[t]
\centering
\hspace{-2em}\includegraphics[width = 1\columnwidth]{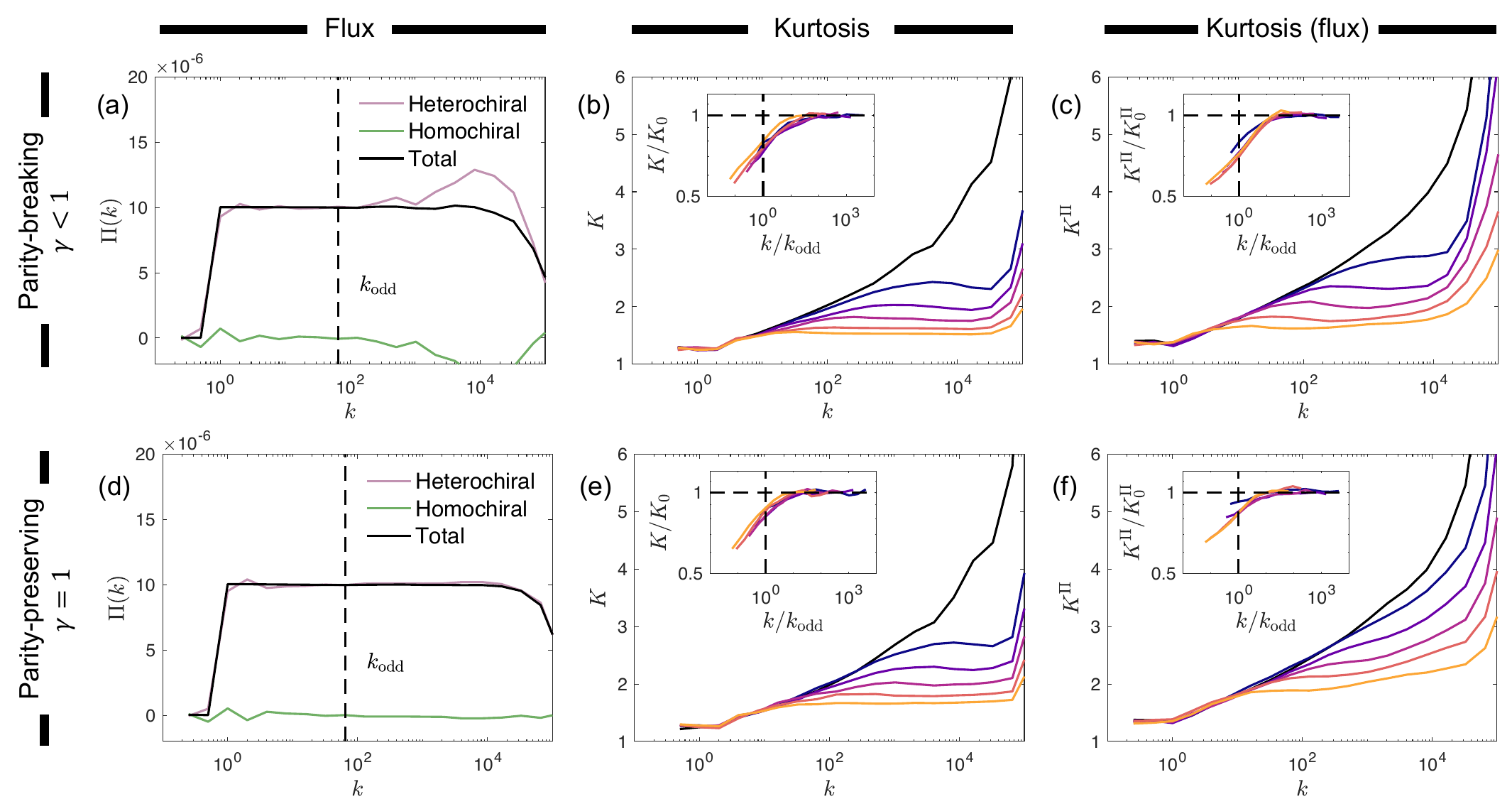}
	\caption{Shell model simulations for wave frequencies $\omega(\bm k)\sim \nu_{\rm odd} k_z|\bm k|$. Inertial range is extended by reducing the normal viscosity, see Sec.~\ref{sec3C}. (a) Energy flux for parity-breaking waves ($\gamma<1$) and $\nu_{\rm odd}=6.6\times 10^{-5}$. (b) Velocity-based kurtosis for parity-breaking waves ($\gamma<1$) and $\nu_{\rm odd}=0,  2.6\times 10^{-7}, 1.0\times 10^{-6}, 4.1\times 10^{-6}, 1.6\times 10^{-5}, 6.6\times 10^{-5}$.. (c) Flux-based kurtosis for parity-breaking waves ($\gamma<1$) and same $\nu_{\rm odd}$ values as in (b). (d) Energy flux for parity-preserving waves ($\gamma=1$) and $\nu_{\rm odd}=6.6\times 10^{-5}$. (e) Velocity-based kurtosis for parity-preserving waves ($\gamma=1$) and $\nu_{\rm odd}$ values same as in (b).  (f) Flux-based kurtosis for parity-preserving waves ($\gamma=1$) and same $\nu_{\rm odd}$ values as in (b). 
	}
	\label{fig.S5}
\end{figure*}

\section{parity-breaking Waves vs parity-preserving Waves}
\label{sec4}
As discussed in Sec.~\ref{sec3}, the multiple scale invariances of the Navier-Stokes equation are broken by the asymmetric effects of (generalized) odd viscosity on homochiral and heterochiral channels. This mechanism only applies to parity-breaking waves which break parity symmetry. For parity-preserving waves, the effects on homochiral and heterochiral channels are symmetric and intermittency may still develop. 

This argument can be tested by the two-channel helical shell model with generalized odd viscosity (Sec.~\ref{sec2B}). The asymmetry between homochiral and heterochiral channels is controlled by the parameter $\gamma$ in Eq.~(\ref{S2.7}): when $\gamma=1$ the two channels are affected in a symmetric way by generalized odd viscosity; when $\gamma<1$ the homochiral channel is less affected. Hence, while the shell model with $\gamma<1$ corresponds to parity-breaking waves with $\omega_{\pm}(\bm k) = \pm c_{\rm odd}k_z|\bm k|^{\alpha-1}$, the shell model with $\gamma=1$ corresponds to parity-preserving waves with, e.g., $\omega_{\pm}(\bm k) =  c_{\rm odd}k_z|\bm k|^{\alpha-1}$. We explore the intermittency for both the parity-breaking and parity-preserving models with $\alpha=1.5$. As expected, the flux-loop is enhanced by $c_{\rm odd}$ in the parity-breaking model (Fig.~\ref{fig.S4}a) but is not enhanced in the parity-preserving model (Fig.~\ref{fig.S4}d). The kurtosis is suppressed by large $c_{\rm odd}$ in the parity-breaking model (Fig.~\ref{fig.S4}b). In the parity-preserving model, on the other hand, the kurtosis still increases with a relatively slow speed for large $c_{\rm odd}$, suggesting that the intermittency is not completely suppressed. This difference is more obvious for the flux-based kurtosis, defined by $K^{\Pi}=S_4^\Pi/S_2^\Pi$, where $S_p^\Pi(k_n) = \langle |\Pi_n/k_n|^{p/3}\rangle$ is the flux-based structure function~\cite{Lvov1998ImprovedSM,de2024extreme}. We find that while the parity-breaking model is still non-intermittent in the wave-affected regime, the parity-preserving model is strongly intermittent (Fig.~\ref{fig.S4}c,f). 

A similar comparison is done for odd waves ($\alpha=2$) (Fig.~\ref{fig.S5}, comparing a,b,c with d,e,f). Surpringly, we find that the velocity-based kurtosis is suppressed in both the parity-breaking and parity-preserving models (Fig.~\ref{fig.S5}b,e). When looking at the flux-based kurtosis we again observe different results: kurtosis does not develop in the parity-breaking model but develops slowly in the parity-preserving model (Fig.~\ref{fig.S5}c,f). This suggests that the parity-breaking model is non-intermittent and the parity-preserving model is intermittent in the wave-affected regime. However, the intermittency of the parity-preserving model is not reflected on the velocity-based kurtosis. This is possibly a coincidence for $\alpha=2$: when $\alpha=2$, the wave-affected solution $E_n\sim\epsilon^{-1/2}\nu_{\rm odd}^{-1/2}k_n^0$ coincides with the equipartition solution (all $u_n$s have similar velocity fluctuations). This leads to strong fluctuations in the wave-affected regime which may mask the non-Gaussian fluctuations of $u_n$. For the 3D Navier-Stokes equation the equipartition solution holds at $E(k)\sim k^2$, hence a similar coincidence may happen at $\alpha=6$ instead of $\alpha=2$. 

\section{Shell model with stochastic waves}
In our two-channel helical shell model, the generalized odd waves in the Navier-Stokes equation are not explicitly included. The effect of generalized odd waves are modelled by the fractions of resonant triads $g_n$. A different approach was adopted in prior shell models of rotating and MHD turbulence, which modifies the classic one-channel shell model with explicit waves~\cite{hattori2004shell,plunian2010cascades}. This approach is summarized in the following form:
\begin{equation}
\begin{aligned}
\partial_t u_{n}=B_n +f_n-\nu k_n^2u_n + i \Omega_n(t)u_n\,,
\end{aligned}
\end{equation}
where 
\begin{equation}
\begin{aligned}
 B_n={a}_n u_{n+2}u_{n+1}^{*}+ b_nu_{n+1}u_{n-1}^{*}-  c_nu_{n-1}u_{n-2}\,
\label{S2.6}
\end{aligned}
\end{equation}
is the nonlinear transfer of Sabra shell model with $a_n=-2b_{n+1}=-2c_{n+2}=k_n$. It has been argued that to reproduce the wave-affected energy spectrum, the wave frequencies $\Omega_n(t)$ cannot be assumed constant, but has to be stochastic with $\langle \Omega_n^2\rangle \sim c^2_{\rm odd} k_n^{2\alpha}$~\cite{hattori2004shell}. The reason for this may be explained by the resonance condition: In the Navier-Stokes equation, generalized odd waves oscillate with frequencies $\omega_\pm(\bm k)=\pm c_{\rm odd}k_z |\bm k|^{\alpha-1}=\pm c_{\rm odd}|\bm k|^{\alpha}\cos(\theta)$, where $\theta$ is the angle between $\bm k$ and $\hat{\bm e}_z$. The 3D distribution of $\theta$ effectively induces a wide distribution of the wave frequencies for given $|\bm k|$, which further guarantees that the resonant condition can be satisfied by some triads. In shell models, however, this 3D structure is absent and the interaction among triads is limited to nearest neighbors. Hence, if constant frequencies $\Omega_n\sim c_{\rm odd} k_n^\alpha$ is assumed, one can show that the resonant condition $\bar{\Omega}_n \tau_{\rm eddy}<1$ will never be satisfied, with $\bar{\Omega}_n=\Omega_{n-1} + \Omega_n + \Omega_{n+1}$. 

The stochasticity of $\Omega_n$ in the shell model effectively reproduces the wide distribution of wave frequencies in the Navier-Stokes equation. The difference is that in the shell model, different $\Omega_n$ is distributed in different time period, while in the Navier-Stokes equation, different frequencies are distributed in different triads. This type of shell models is able to reproduce the average energy transfer of the Navier-Stokes equation, however, it would be inappropriate to study intermittency with these models. The reason is that the stochasticity naturally breaks scale invariances at any given time, hence intermittency must be suppressed. This suppression is just caused by an artifact of the stochastic shell model that is absent in the Navier-Stokes equation.

\bibliography{citation}